\documentclass[12pt]{article}
\def\ap{a^{\dagger}}
\def\alg{${\cal A}^{(\lambda)}_{\alpha_0 \alpha_1 \ldots \alpha_{\lambda-2}}$}
\def\algtwo{${\cal A}^{(2)}_{\alpha_0}$}
\def\algthree{${\cal A}^{(3)}_{\alpha_0 \alpha_1}$}
\def\algthreebis{${\cal A}^{(3)}_{\alpha_1, - \alpha_0 - \alpha_1}$}
\def\algthreeter{${\cal A}^{(3)}_{- \alpha_0 - \alpha_1, \alpha_0}$}
\def\case#1#2{{\textstyle{#1\over #2}}}
\def\half{\case{1}{2}}
\def\RE{\mathop{\Re e}\nolimits}
\def\IM{\mathop{\Im m}\nolimits}
\def\Ap{A^{\dagger}}
\def\Qp{Q^{\dagger}}

\newcommand{\mod}{\mathop{\rm mod}\nolimits}

\newcommand{\level}{\thicklines\line(1,0){10}}
\newcommand{\trait}{\line(0,1){1}}
\newcommand{\ord}{\thicklines\line(1,0){2}}
\newcommand{\separation}{\line(0,1){87.5}}
\newsavebox{\two}
\newsavebox{\three}
\newsavebox{\four}
\newsavebox{\five}
\newsavebox{\six}
\newsavebox{\dash}
\savebox{\two}(10,15){\multiput(0,0)(0,15){2}{\level}}
\savebox{\three}(10,30){\multiput(0,0)(0,15){3}{\level}}
\savebox{\four}(10,45){\multiput(0,0)(0,15){4}{\level}}
\savebox{\five}(10,60){\multiput(0,0)(0,15){5}{\level}}
\savebox{\six}(10,75){\multiput(0,0)(0,15){6}{\level}}
\savebox{\dash}(5,10)[b]{\multiput(0,0)(0,2){3}{\trait}}
\title{
\hfill{\normalsize ULB/229/CQ/98/1}\\
\vspace{1cm}
Algebraic Realization of Supersymmetric Quantum Mechanics for Cyclic Shape
Invariant Potentials}
\author{C. Quesne \thanks{Directeur de recherches FNRS; E-mail:
cquesne@ulb.ac.be}\ , N. Vansteenkiste \thanks{E-mail: nvsteen@ulb.ac.be}\\
{\small\sl Physique Nucl\'eaire Th\'eorique et Physique Math\'ematique,
Universit\'e Libre de Bruxelles,}\\
{\small\sl Campus de la Plaine CP229, Boulevard du
Triomphe, B-1050 Brussels, Belgium}}
\date{}
\begin{document}
\maketitle
\begin{abstract}
We study in detail the spectrum of the bosonic oscillator Hamiltonian
associated with the $C_3$-extended oscillator algebra \algthree, where $C_3$
denotes a cyclic group of order~three, and classify the various types of
spectra in
terms of the algebra parameters $\alpha_0$,~$\alpha_1$. In such a
classification,
we identify those spectra having an infinite number of periodically spaced
levels,
similar to those of cyclic shape invariant potentials of period three. We
prove that
the hierarchy of supersymmetric Hamiltonians and supercharges, corresponding to
the latter, can be realized in terms of some appropriately chosen \algthree\
algebras, and of Pauli spin matrices. Extension to period-$\lambda$ spectra in
terms of $C_{\lambda}$-extended oscillator algebras is outlined.
\end{abstract}
%
%
\section{Introduction}  \label{sec:intro}
When supplemented with the concept of shape invariance~\cite{gendenshtein},
supersymmetric quantum mechanics (SSQM)~\cite{witten} has proved very useful
for generating exactly solvable quantum mechanical models. Devising new
approaches to construct shape invariant potentials is still under current
investigation (for a recent review see Ref.~\cite{cooper}). A recent
advance in this
field has been the introduction of cyclic shape invariant potentials by
Sukhatme {\it et al\/}~\cite{sukhatme}, generalizing a previous work of
Gangopadhyaya and Sukhatme~\cite{gango}.\par
%
%
In addition, SSQM has established a nice symmetry between bosons and
fermions~\cite{witten}. Such a symmetry has been extended to some exotic
statistics. Replacing fermions by parafermions~\cite{green},
pseudofermions~\cite{beckers95}, or orthofermions~\cite{mishra}, for instance,
has led to parasupersymmetric (PSSQM)~\cite{rubakov,beckers90},
pseudosupersymmetric~\cite{beckers95}, or orthosupersymmetric~\cite{khare}
quantum mechanics, respectively.\par
%
%
The development of quantum groups and quantum algebras~\cite{drinfeld} during
the last decade has proved very useful in connection with such problems. In
particular, various deformations and extensions of the oscillator algebra have
found a lot of applications to quantum mechanics, in general, and to SSQM
and some
of its generalizations, in particular.\par
%
%
Deformations of the oscillator algebra arose from successive generalizations of
the Arik-Coon~\cite{arik}, and Biedenharn-Macfarlane~\cite{biedenharn}
$q$-oscillators. Various attempts have been made to introduce some order in the
various deformations by defining `generalized deformed oscillator algebras'
(GDOAs). Among them, one may quote the treatments due to Jannussis {\it et
al\/}~\cite{jannussis}, Daskaloyannis~\cite{daska91}, Irac-Astaud and
Rideau~\cite{irac}, McDermott and Solomon~\cite{mcdermott}, Meljanac {\it et
al\/}~\cite{meljanac}, Katriel and Quesne~\cite{katriel}, Quesne and
Vansteenkiste~\cite{cq95b, cq96}. In the remainder of the present paper, we
shall
refer to GDOAs as defined in Ref.~\cite{cq95b}. GDOAs have found some
interesting applications to the algebraic treatment of some one-dimensional
exactly solvable potentials~\cite{daska92, cq94a} or two-dimensional
superintegrable systems~\cite{bonatsos93a}, as well as to the description of
systems with non-standard statistics~\cite{meljanac,greenberg, chaturvedi,
cq94b}.\par
%
%
$G$-extended oscillator (or alternatively Heisenberg\footnote{In both the
oscillator and Heisenberg algebras, the creation and annihilation operators
$\ap$,
$a$ are considered as generators, but in the former the number operator
appears as
an additional independent generator, whereas in the latter it is defined in
terms of
$\ap$, $a$ as $N \equiv \ap a$.}) algebras, where $G$ is some finite group,
essentially appeared in connection with $n$-particle integrable models. It was
shown that they provide an algebraic formulation~\cite{vasiliev, poly,
brze} of the
Calogero model~\cite{calogero} or some generalizations thereof~\cite{cq95a}. In
the former case, $G$ is the symmetric group $S_n$~\cite{poly}. For two
particles,
the abelian group~$S_2$ can be realized in terms of Klein operator $K = (-1)^N$,
where $N$ denotes the number operator. The $S_2$-extended oscillator algebra is
then known as the Calogero-Vasiliev~\cite{vasiliev}, or modified~\cite{brze}
oscillator algebra.\par
%
%
The usefulness of GDOAs in connection with SSQM was pointed out by Bonatsos and
Daskaloyannis~\cite{bonatsos93b}. Then Brzezi\'nski {\it et al\/}~\cite{brze}, and
Plyushchay~\cite{plyu} in more detail (see also Ref.~\cite{beckers97}), showed
that the Calogero-Vasiliev algebra provides a minimal bosonization of SSQM in
terms of boson-like particles, instead of a combination of bosons and
fermions, as
is the case in the standard approach~\cite{witten}.\par
%
%
In a recent work~\cite{cq98a}, we introduced a new type of $G$-extended
oscillator
algebras\linebreak \alg, where $G$ is a cyclic group of order~$\lambda$,
$C_{\lambda} = \{\, I, T, T^2, \ldots, T^{\lambda-1} \,\}$, and $\alpha_0$,
$\alpha_1$, $\ldots$,~$\alpha_{\lambda-2}$ denote $\lambda-1$ independent real
parameters. Since $C_{\lambda}$ is an Abelian group, its elements can be
realized
in terms of~$N$ only, so that \alg\ becomes a GDOA. The cyclic group~$C_2$ being
isomorphic to~$S_2$, the $C_2$-extended oscillator algebra~\algtwo\ is
equivalent to  Calogero-Vasiliev algebra. Hence, new features only appear
for~$\lambda \ge 3$.\par
%
%
To each \alg\ algebra, one can associate a bosonic oscillator Hamiltonian~$H_0$.
That corresponding to \algtwo\ is just the two-particle Calogero Hamiltonian,
which has a very simple spectrum, coinciding with that of a shifted harmonic
oscillator. For higher $\lambda$ values, the situation is entirely different as,
according to the parameter values, the spectrum may be nondegenerate, or may
exhibit some $(\nu + 1)$-fold degeneracies, where $\nu$ may take any value
in the
set $\{1, 2, \ldots, \lambda-1\}$, with in each case various possibilities
for the
level ordering.\par
%
%
In \cite{cq98a}, we extended Plyushchay's work by showing that the
$C_3$-extended oscillator algebra \algthree\ provides a minimal bosonization of
Rubakov-Spiridonov PSSQM of order $p=2$~\cite{rubakov}. More generally, it
can be
proved~\cite{cq98b} that \alg\ leads to the same result for Rubakov-Spiridonov
PSSQM of order~$p = \lambda -1$.\par
%
%
Here, we will address the problem of SSQM for cyclic shape invariant
potentials of
period~$\lambda$. We will prove that the corresponding hierarchy of
supersymmetric Hamiltonians and supercharges, which repeats after a cycle of
$\lambda$~iterations can be realized in terms of some appropriate \alg\
algebras,
and of Pauli spin matrices. Although the detailed derivation will be
carried out for
the simplest nontrivial case corresponding to $\lambda=3$, it will become clear
that the arguments are still valid for arbitrary $\lambda \ge 3$.\par
%
%
To deal with this problem, after reviewing the definitions of the $C_3$-extended
oscillator algebra, and of the corresponding oscillator Hamiltonian in
section~\ref{sec:Hamiltonian}, we will study in detail the $H_0$ spectrum
associated with \algthree, and derive the complete classification of the
different
types of spectra in terms of the algebra parameters $\alpha_0$, $\alpha_1$, in
section~\ref{sec:spectra}. In section~\ref{sec:periodic} , we will then identify
those spectra having an infinite number of periodically spaced levels, and show
that for some of them one can obtain the searched for algebraic realization of
SSQM. Section~\ref{sec:conclusion} contains some concluding
remarks about the extension to period-$\lambda$ spectra.\par
%
%
\section{$C_3$-Extended Oscillator Algebra and Hamiltonian}
\label{sec:Hamiltonian}
\hspace{\parindent}
Let us consider the bosonic oscillator Hamiltonian, defined (in units wherein
$\hbar \omega = 1$) by~\cite{cq98a}
\begin{equation}
  H_0 \equiv \case{1}{2} \left\{a, \ap\right\},   \label{eq:H}
\end{equation}
where the creation and annihilation operators $\ap$, $a$ satisfy the generalized
relations
\begin{eqnarray}
  \left[N, \ap\right] & = & \ap, \qquad [N, T] = 0, \qquad T^3 = I, \nonumber \\
  \left[a, \ap\right] & = & I + \kappa_1 T + \kappa_2 T^2,
            \qquad \ap T = e^{-2\pi i/3}\, T \ap,   \label{eq:C3-com}
\end{eqnarray}
together with their Hermitian conjugates. Here, $N = N^{\dagger}$ is the number
operator, $T = \left(T^{\dagger}\right)^{-1}$ is the (unitary) generator of
a cyclic
group $C_3 = \{I, T, T^2\}$, and $\kappa_1$, $\kappa_2$ are two complex
constants,
restricted by the condition $\kappa_2 = \kappa_1^*$ (deriving from the relation
$T^2 = T^{\dagger}$).\par
%
%
In the present paper, we shall be concerned with a realization of~$T$ as a
function
of~$N$, given by
\begin{equation}
  T = e^{2\pi iN/3},
\end{equation}
in which case there only remain two nontrivial relations in
equation~(\ref{eq:C3-com}), namely
\begin{equation}
  \left[N, \ap\right] = \ap, \qquad  \left[a, \ap\right] = I + 2 (\RE\kappa_1)
  \cos\case{2\pi}{3}N - 2 (\IM\kappa_1) \sin\case{2\pi}{3}N.
  \label{eq:GDOA-com}
\end{equation}
According to \cite{cq95b}, equation~(\ref{eq:GDOA-com}) defines a GDOA
${\cal A}(G(N))$, with
\begin{equation}
  G(N) \equiv I + 2 (\RE\kappa_1) \cos\case{2\pi}{3}N - 2 (\IM\kappa_1)
 \sin\case{2\pi}{3}N.
\end{equation}
\par
%
%
Provided its parameters satisfy some conditions to be given below, the algebra
possesses a bosonic Fock space ${\cal F} = \{\, |n\rangle \mid n = 0, 1, 2,
\ldots
\,\}$, spanned by the normalized eigenvectors of $N$,
\begin{equation}
  N |n\rangle = n |n\rangle, \qquad \langle n | m \rangle = \delta_{n,m},
\end{equation}
which can be written as
\begin{equation}
  |n\rangle = {\cal N}_n^{-1/2} \left(\ap\right)^n |0\rangle, \qquad n = 0,
1, 2,
  \ldots,    \label{eq:n-state}
\end{equation}
where ${\cal N}_n$ is some normalization constant, and $|0\rangle$ is a vacuum
state, i.e.,
\begin{equation}
  a |0\rangle = 0.
\end{equation}
\par
%
%
{}From equation~(\ref{eq:GDOA-com}), it is clear that the operators $\ap$,
$a$ act
differently in the three subspaces ${\cal F}_{\mu}$, $\mu = 0$, 1,~2,
of~$\cal F$,
defined by ${\cal F}_{\mu} \equiv \{\, |3k + \mu\rangle \mid k = 0, 1, 2,
\ldots\,\}$,
and such that ${\cal F} = {\cal F}_0 \oplus {\cal F}_1 \oplus {\cal F}_2$.
Actually,
these three subspaces are the carrier spaces of the three inequivalent
irreducible
(one-dimensional) matrix representations of~$C_3$, defined by $\Gamma^{\mu}(T)
= \exp(2\pi i \mu /3)$, $\mu = 0$, 1, 2~\cite{cornwell}. The projection
operators
$P_{\mu}$ on the ${\cal F}_{\mu}$ subspaces are given by $P_{\mu} = \frac{1}{3}
\sum_{\nu=0}^2  \exp(-2\pi i\mu\nu/3)\, T^{\nu}$, or
\begin{eqnarray}
  P_0 & = & \case{1}{3} \left(I + 2 \cos\case{2\pi}{3}N\right), \qquad
           P_1 = \case{1}{3} \left(I - \cos\case{2\pi}{3}N + \sqrt{3}
           \sin\case{2\pi}{3}N \right), \nonumber \\
  P_2 & = & \case{1}{3} \left(I - \cos\case{2\pi}{3}N - \sqrt{3}
\sin\case{2\pi}{3}N
           \right).    \label{eq:proj}
\end{eqnarray}
As it can easily be checked on equation~(\ref{eq:proj}), the $P_{\mu}$'s
satisfy the
relations
\begin{equation}
  P_{\mu} P_{\nu} = \delta_{\mu,\nu} P_{\mu}, \qquad \sum_{\mu=0}^2 P_{\mu} = I,
\end{equation}
as it should be.\par
%
%
In terms of such operators, equation~(\ref{eq:GDOA-com}) can be rewritten as
\begin{equation}
  \left[N, \ap\right] = \ap, \qquad  \left[a, \ap\right] = I + \alpha_0 P_0
+ \alpha_1
  P_1 + \alpha_2 P_2,      \label{eq:GDOA-combis}
\end{equation}
where $\alpha_{\mu}$, $\mu=0$, 1,~2, are three real parameters, connected with
$\kappa_1$ and $\kappa_2 = \kappa_1^*$ by the relations $\alpha_{\mu} =
\sum_{\nu=1}^2 \exp(2\pi i\mu\nu/3)\, \kappa_{\nu}$, or
\begin{equation}
  \alpha_0 = 2 \RE\kappa_1, \quad \alpha_1 = - \RE\kappa_1 - \sqrt{3}
  \IM\kappa_1, \quad \alpha_2 = - \alpha_0 - \alpha_1 = - \RE\kappa_1 + \sqrt{3}
  \IM\kappa_1.
\end{equation}
Hence, we may also express $G(N)$ as
\begin{equation}
  G(N) = I + \alpha_0 P_0 + \alpha_1 P_1 + \alpha_2 P_2, \qquad \mbox{with\ }
  \alpha_0 + \alpha_1 + \alpha_2 = 0,
\end{equation}
and denote the algebra ${\cal A}(G(N))$ by \algthree, where the two independent
real parameters $\alpha_0$, $\alpha_1$ are specified. In the remainder of this
paper, we will assume $\alpha_{\mu} \equiv \alpha_{\mu\mod3}$, and
$P_{\mu} \equiv P_{\mu\mod3}$ for arbitrary integer $\mu$ values. \par
%
%
{}For any GDOA, one may define a so-called structure function $F(N)$, which
is the
solution of the difference equation $F(N+1) - F(N) = G(N)$, such that $F(0) =
0$~\cite{daska91, katriel, cq95b, cq96, bonatsos93b}. In the present case,
we get
\begin{equation}
  F(N) = N + \beta_1 P_1 + \beta_2 P_2, \qquad \mbox{where\ $\beta_1 \equiv
  \alpha_0$, $\beta_2 \equiv \alpha_0 + \alpha_1$}.   \label{eq:F}
\end{equation}
\par
%
%
In the bosonic Fock space~$\cal F$, $F(N)$ satisfies the relations
\begin{equation}
  \ap a = F(N), \qquad a \ap = F(N+1),   \label{eq:F-Fock}
\end{equation}
and the normalization coefficient ${\cal N}_n$ in equation~(\ref{eq:n-state}) is
given by ${\cal N}_n = \prod_{i=1}^n F(i)$, or
\begin{eqnarray}
  {\cal N}_{3k} & = & 3^{3k} \left[\Gamma(\overline{\beta}_1)
           \Gamma(\overline{\beta}_2)\right]^{-1} \Gamma(k+1) \Gamma(k +
           \overline{\beta}_1) \Gamma(k + \overline{\beta}_2), \nonumber \\
  {\cal N}_{3k+1} & = & 3^{3k+1} \left[\Gamma(\overline{\beta}_1)
           \Gamma(\overline{\beta}_2)\right]^{-1} \Gamma(k+1) \Gamma(k + 1 +
           \overline{\beta}_1) \Gamma(k + \overline{\beta}_2), \\
  {\cal N}_{3k+2} & = & 3^{3k+2} \left[\Gamma(\overline{\beta}_1)
           \Gamma(\overline{\beta}_2)\right]^{-1} \Gamma(k+1) \Gamma(k + 1 +
           \overline{\beta}_1) \Gamma(k + 1 + \overline{\beta}_2), \nonumber
\end{eqnarray}
in terms of gamma functions, and of $\overline{\beta}_1 \equiv (\beta_1 + 1)/3$,
$\overline{\beta}_2 \equiv (\beta_2 + 2)/3$. The creation and annihilation
operators
act upon~$|n\rangle$ as
\begin{equation}
  \ap |n\rangle = \sqrt{F(n+1)}\, |n+1\rangle, \qquad a |n\rangle =
\sqrt{F(n)}\,
  |n-1\rangle.
\end{equation}
Hence, from equation~(\ref{eq:F}), it is obvious that $\cal F$ exists if
and only if
$F(1) > 0$ and $F(2) > 0$, or, in other words, the algebra parameters are
restricted
to those values for which
\begin{equation}
  \alpha_0 > - 1, \qquad \alpha_1 > - 2 - \alpha_0.     \label{eq:domain}
\end{equation}
We shall henceforth assume that these conditions are fulfilled. Note that
$\alpha_0
= \alpha_1 = 0$ corresponds to the standard harmonic oscillator.\par
%
%
It is now straightforward to determine the action of the bosonic oscillator
Hamiltonian~$H_0$, defined in equation~(\ref{eq:H}), in the bosonic Fock
space~$\cal F$. For such a purpose, it is useful to rewrite $H_0$ in the
equivalent
forms
\begin{equation}
  H_0 = \ap a + \case{1}{2} \left(I + \alpha_0 P_0 + \alpha_1 P_1 + \alpha_2 P_2
  \right) = N + \case{1}{2} I + \gamma_0 P_0 + \gamma_1 P_1 + \gamma_2 P_2
  \label{eq:Hbis}
\end{equation}
by using equations~(\ref{eq:GDOA-combis}), (\ref{eq:F}),
and~(\ref{eq:F-Fock}). In
equation~(\ref{eq:Hbis}), the parameters $\gamma_{\mu}$, $\mu = 0$, 1,~2, are
defined by
\begin{equation}
  \gamma_0 \equiv \case{1}{2} \alpha_0, \qquad \gamma_1 \equiv \case{1}{2}
  (2\alpha_0 + \alpha_1), \qquad \gamma_2 \equiv \case{1}{2} (\alpha_0 +
  \alpha_1),    \label{eq:gamma}
\end{equation}
and satisfy the relation $\gamma_0 - \gamma_1 + \gamma_2 = 0$. The number
operator eigenvectors $|n\rangle = |3k + \mu\rangle$ are also eigenvectors
of $H_0$, corresponding to the eigenvalues
\begin{equation}
  E_{3k+\mu} = 3k + \mu + \gamma_{\mu} + \case{1}{2}, \qquad k = 0,
 1, 2, \ldots, \qquad \mu = 0, 1, 2.    \label{eq:energy}
\end{equation}
\par
%
%
In each ${\cal F}_{\mu}$ subspace of~$\cal F$, the spectrum of~$H_0$ is
therefore
harmonic, but the three infinite sets of equally spaced energy levels,
corresponding
to $\mu = 0$, 1,~2, respectively, may be shifted with respect to each other
by some
amounts depending upon the algebra parameters $\alpha_0$,~$\alpha_1$ through
their linear combinations $\gamma_0$, $\gamma_1$,~$\gamma_2$, defined in
equation~(\ref{eq:gamma}). We may therefore obtain nondegenerate spectra,
as well
as spectra with some double or triple degeneracies. In the next section, we will
study such spectra in detail.\par
%
%
\section{Classification of $C_3$-Extended Oscillator Hamiltonian Spectra}
\label{sec:spectra}
\setcounter{equation}{0}
\hspace{\parindent}
To obtain the various types of $H_0$ spectra, we shall proceed in two steps. We
shall first determine the possible orderings of the $H_0$ ground states in
${\cal
F}_0$, ${\cal F}_1$, and~${\cal F}_2$, corresponding to the eigenvalues $E_0$,
$E_1$, and~$E_2$, respectively. This will give rise to three general and two
intermediate classes of spectra. Then, for each of these five possibilities, we
shall successively study the relative order of the excited states in ${\cal
F}_0$,
${\cal F}_1$, and~${\cal F}_2$ in the nondegenerate, doubly- and
triply-degenerate cases.\par
%
%
Considering first $E_0$, $E_1$, and~$E_2$, we obtain from
equations~(\ref{eq:gamma}) and (\ref{eq:energy})
\begin{equation}
  E_1 - E_0 = \half (\alpha_0 + \alpha_1 + 2), \qquad E_2 - E_1 = \half (2 -
  \alpha_0), \qquad E_2 - E_0 = \half (\alpha_1 + 4).
\end{equation}
Since the parameter values are restricted by equation~(\ref{eq:domain}), it is
obvious that the ground states in ${\cal F}_0$, ${\cal F}_1$, and~${\cal
F}_2$ may
either be nondegenerate, or exhibit a double degeneracy. In the former
case, they
may be ordered in three different ways, which we will refer to as (I),
(II), and~(III),
respectively, as listed hereafter
\begin{eqnarray}
     & \mbox{(I)} & E_0 < E_1 < E_2 \qquad \mbox{if $-1 < \alpha_0 < 2$ and
              $-2-\alpha_0 < \alpha_1$}, \nonumber \\
     & \mbox{(II)} & E_0 < E_2 < E_1 \qquad \mbox{if $2 < \alpha_0$ and
              $-4 < \alpha_1$}, \label{eq:general} \\
     & \mbox{(III)} & E_2 < E_0 < E_1 \qquad \mbox{if $2 < \alpha_0$ and
              $-2-\alpha_0 < \alpha_1 < -4$}. \nonumber
\end{eqnarray}
In the latter case, their ordering is intermediate between classes (I)
and~(II), or
(II) and~(III), and are given by
\begin{eqnarray}
   & \mbox{(I-II)} & E_0 < E_1 = E_2 \qquad \mbox{if $\alpha_0 = 2$ and
              $-4 < \alpha_1$}, \nonumber \\
   & \mbox{(II-III)} & E_0 = E_2 < E_1 \qquad \mbox{if $2 < \alpha_0$ and
              $\alpha_1 = -4$},   \label{eq:interm}
\end{eqnarray}
respectively.\par
%
%
Let us now consider the excited states in ${\cal F}_0$, ${\cal F}_1$, and~${\cal
F}_2$, and distinguish between nondegenerate, doubly- and triply-degenerate
spectra.\par
%
%
\subsection{Nondegenerate spectra}   \label{sec:nondeg}
\hspace{\parindent}
{}For nondegenerate spectra, we have only to consider the three general
classes (I),
(II), and~(III).\par
%
%
Starting with class~(I), we note that since $E_3 -E_2 = (2 - \alpha_1)/2$,
and $E_3
- E_1 = (4 - \alpha_0 - \alpha_1)/2$, we have three different possibilities
for the
ordering of~$E_3$ with respect to $E_1$, and~$E_2$:
\begin{eqnarray}
     & E_0 < E_1 < E_2 < E_3 \quad & \mbox{if $-1 < \alpha_0 < 2$ and
              $-2-\alpha_0 < \alpha_1 < 2$}, \nonumber \\
     & E_0 < E_1 < E_3 < E_2 \quad & \mbox{if $-1 < \alpha_0 < 2$ and
              $2 < \alpha_1 < 4 - \alpha_0$}, \\
     & E_0 < E_3 < E_1 < E_2 \quad & \mbox{if $-1 < \alpha_0 < 2$ and
              $4-\alpha_0 < \alpha_1$}. \nonumber
\end{eqnarray}
Furthermore, since $E_4 - E_2 = (\alpha_0 + 4)/2$ is positive over the whole
parameter range, in the first two cases the remainder of the spectrum is
entirely
determined, so that we obtain $E_0 < E_1 < E_2 < E_3 < E_4 < E_5 < E_6 <
\cdots$,
and $E_0 < E_1 < E_3 < E_2 < E_4 < E_6 < E_5 < \cdots$, respectively.\par
%
%
In the third case, we have to study the ordering of~$E_6$ with respect to $E_1$,
and~$E_2$. As $E_6 -E_2 = (8 - \alpha_1)/2$, and $E_6 - E_1 = (10 - \alpha_0 -
\alpha_1)/2$, there again appear three different possibilities:
\begin{eqnarray}
     & E_0 < E_3 < E_1 < E_2 < E_6 \quad & \mbox{if $-1 < \alpha_0 < 2$ and
              $4-\alpha_0 < \alpha_1 < 8$}, \nonumber  \\
     & E_0 < E_3 < E_1 < E_6 < E_2 \quad & \mbox{if $-1 < \alpha_0 < 2$ and
              $8 < \alpha_1 < 10 - \alpha_0$}, \\
     & E_0 < E_3 < E_6 < E_1 < E_2 \quad & \mbox{if $-1 < \alpha_0 < 2$ and
              $10 - \alpha_0 < \alpha_1$}, \nonumber
\end{eqnarray}
where for the first two the remainder of the spectrum is entirely
determined.\par
%
%
By recursively carrying on such a classification, we get two nondegenerate
spectra
subclasses (I.1) and (I.2), themselves labelled by some index~$n$ running
over 1, 2,
3,~$\ldots$:
\begin{eqnarray}
  & \mbox{(I.1.$n$)} & E_0 < E_3 < \cdots < E_{3n-3} < E_1 < E_2 < E_{3n} <
E_4 <
            E_5 < \cdots \nonumber \\
  & & \mbox{if $-1 < \alpha_0 < 2$ and $6n - \alpha_0 - 8 < \alpha_1 < 6n - 4$},
            \nonumber \\
  & \mbox{(I.2.$n$)} & E_0 < E_3 < \cdots < E_{3n-3} < E_1 < E_{3n} < E_2 <
E_4 <
            E_{3n+3} \label{eq:I-nondeg} \\
  & & < E_5 < \cdots \nonumber \\
  & & \mbox{if $-1 < \alpha_0 < 2$ and $ 6n - 4 < \alpha_1 < 6n - \alpha_0
- 2$}.
            \nonumber
\end{eqnarray}
The parameter values in equation~(\ref{eq:I-nondeg}) can simply be obtained by
combining those defining class~(I) in equation~(\ref{eq:general}) with the
conditions $E_{3n-3} - E_1 = (6n - \alpha_0 - \alpha_1 - 8)/2 < 0$ for both
subclasses, and either $E_{3n} - E_2 = (6n - \alpha_1 - 4)/2 > 0$ for the
first one,
or $E_{3n} - E_1 = (6n - \alpha_0 - \alpha_1 - 2)/2 > 0$ and $E_{3n} - E_2
= (6n -
\alpha_1 - 4)/2 < 0$ for the second one.\par
%
%
It is worth noting that the parameter values corresponding to type (I.1.$n$) and
(I.2.$n$) spectra cover all class~(I) parameter range, but for $-1 <
\alpha_0 < 2$,
$\alpha_1 = 6n - 4$ or $\alpha_1 = 6n - \alpha_0 - 2$, where $n = 1$, 2,
3,~$\ldots$.\par
%
%
\begin{figure}

\begin{picture}(130,95)(-37,-10)

\put(-10,-2.5){\thicklines\vector(0,1){87.5}}
\multiput(-11,0)(0,15){6}{\ord}
\put(-17,0){\makebox(5,0)[r]{1/2}}
\put(-17,15){\makebox(5,0)[r]{7/2}}
\put(-17,30){\makebox(5,0)[r]{13/2}}
\put(-17,45){\makebox(5,0)[r]{19/2}}
\put(-17,60){\makebox(5,0)[r]{25/2}}
\put(-17,75){\makebox(5,0)[r]{31/2}}

\put(50,-2.5){\separation}

\put(20,-7.5){\makebox(0,0){(a)}}

\put(0,0){\usebox{\six}}
\put(5,-2){\makebox(0,0){0}}
\put(5,13){\makebox(0,0){3}}
\put(5,28){\makebox(0,0){6}}
\put(5,43){\makebox(0,0){9}}
\put(5,58){\makebox(0,0){12}}
\put(5,73){\makebox(0,0){15}}
\put(5,77){\usebox{\dash}}

\put(15,20){\usebox{\four}}
\put(20,18){\makebox(0,0){1}}
\put(20,33){\makebox(0,0){4}}
\put(20,48){\makebox(0,0){7}}
\put(20,63){\makebox(0,0){10}}
\put(20,67){\usebox{\dash}}

\put(30,25){\usebox{\four}}
\put(35,23){\makebox(0,0){2}}
\put(35,38){\makebox(0,0){5}}
\put(35,53){\makebox(0,0){8}}
\put(35,68){\makebox(0,0){11}}
\put(35,72){\usebox{\dash}}


\put(80,-7.5){\makebox(0,0){(b)}}

\put(60,0){\usebox{\six}}
\put(65,-2){\makebox(0,0){0}}
\put(65,13){\makebox(0,0){3}}
\put(65,28){\makebox(0,0){6}}
\put(65,43){\makebox(0,0){9}}
\put(65,58){\makebox(0,0){12}}
\put(65,73){\makebox(0,0){15}}
\put(65,77){\usebox{\dash}}

\put(75,27.5){\usebox{\four}}
\put(80,25.5){\makebox(0,0){1}}
\put(80,40.5){\makebox(0,0){4}}
\put(80,55.5){\makebox(0,0){7}}
\put(80,70.5){\makebox(0,0){10}}
\put(80,74.5){\usebox{\dash}}

\put(90,32.5){\usebox{\four}}
\put(95,30.5){\makebox(0,0){2}}
\put(95,45.5){\makebox(0,0){5}}
\put(95,60.5){\makebox(0,0){8}}
\put(95,75.5){\makebox(0,0){11}}
\put(95,79.5){\usebox{\dash}}

\end{picture}

\caption{Examples of nondegenerate $H_0$~spectra belonging to class~(I): (a)
type (I.1.2) spectrum with $\alpha_0 = 0$, $\alpha_1 = 6$; (b) type (I.2.2)
spectrum
with $\alpha_0 = 0$, $\alpha_1 = 9$.}

\end{figure}
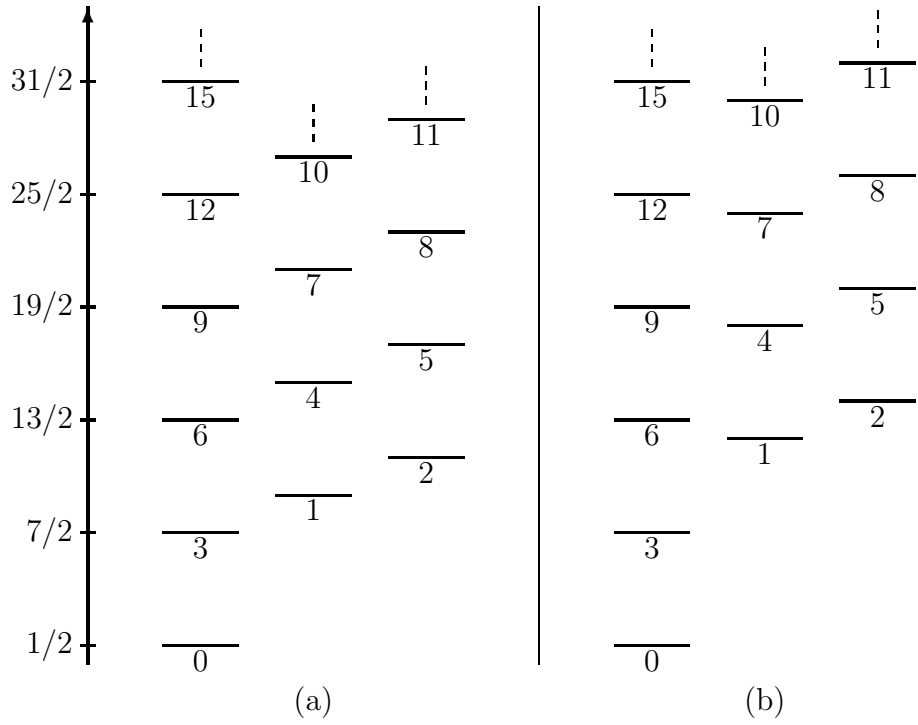
%
%
\begin{figure}

\begin{picture}(130,95)(-37,-10)

\put(-10,-2.5){\thicklines\vector(0,1){87.5}}
\multiput(-11,0)(0,15){6}{\ord}
\put(-17,0){\makebox(5,0)[r]{11/2}}
\put(-17,15){\makebox(5,0)[r]{17/2}}
\put(-17,30){\makebox(5,0)[r]{23/2}}
\put(-17,45){\makebox(5,0)[r]{29/2}}
\put(-17,60){\makebox(5,0)[r]{35/2}}
\put(-17,75){\makebox(5,0)[r]{41/2}}

\put(50,-2.5){\separation}

\put(20,-7.5){\makebox(0,0){(a)}}

\put(0,0){\usebox{\six}}
\put(5,-2){\makebox(0,0){0}}
\put(5,13){\makebox(0,0){3}}
\put(5,28){\makebox(0,0){6}}
\put(5,43){\makebox(0,0){9}}
\put(5,58){\makebox(0,0){12}}
\put(5,73){\makebox(0,0){15}}
\put(5,77){\usebox{\dash}}

\put(15,40){\usebox{\three}}
\put(20,38){\makebox(0,0){1}}
\put(20,53){\makebox(0,0){4}}
\put(20,68){\makebox(0,0){7}}
\put(20,72){\usebox{\dash}}

\put(30,20){\usebox{\four}}
\put(35,18){\makebox(0,0){2}}
\put(35,33){\makebox(0,0){5}}
\put(35,48){\makebox(0,0){8}}
\put(35,63){\makebox(0,0){11}}
\put(35,67){\usebox{\dash}}


\put(80,-7.5){\makebox(0,0){(b)}}

\put(60,0){\usebox{\six}}
\put(65,-2){\makebox(0,0){0}}
\put(65,13){\makebox(0,0){3}}
\put(65,28){\makebox(0,0){6}}
\put(65,43){\makebox(0,0){9}}
\put(65,58){\makebox(0,0){12}}
\put(65,73){\makebox(0,0){15}}
\put(65,77){\usebox{\dash}}

\put(75,47.5){\usebox{\three}}
\put(80,45.5){\makebox(0,0){1}}
\put(80,60.5){\makebox(0,0){4}}
\put(80,75.5){\makebox(0,0){7}}
\put(80,79.5){\usebox{\dash}}

\put(90,27.5){\usebox{\four}}
\put(95,25.5){\makebox(0,0){2}}
\put(95,40.5){\makebox(0,0){5}}
\put(95,55.5){\makebox(0,0){8}}
\put(95,70.5){\makebox(0,0){11}}
\put(95,74.5){\usebox{\dash}}

\end{picture}

\caption{Examples of nondegenerate $H_0$~spectra belonging to class~(II): (a)
type (II.1.2.2) spectrum with $\alpha_0 = 10$, $\alpha_1 = 4$; (b) type
(II.2.2.2)
spectrum with $\alpha_0 = 10$, $\alpha_1 = 7$.}

\end{figure}
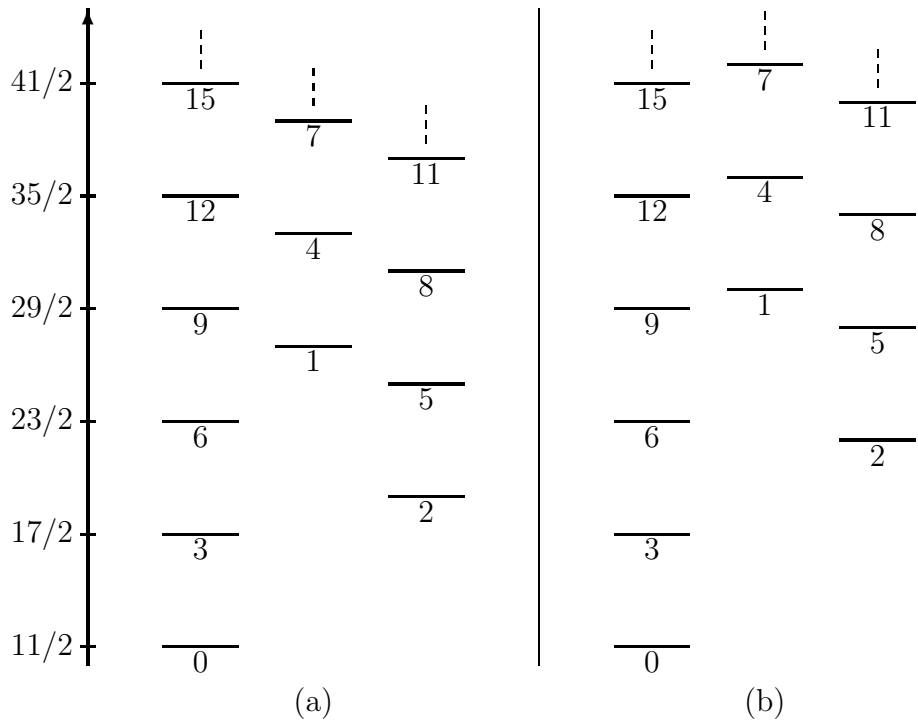
%
%
%
\begin{figure}

\begin{picture}(130,95)(-37,-10)

\put(-10,-2.5){\thicklines\vector(0,1){87.5}}
\multiput(-11,0)(0,15){6}{\ord}
\put(-17,0){\makebox(5,0)[r]{11/2}}
\put(-17,15){\makebox(5,0)[r]{17/2}}
\put(-17,30){\makebox(5,0)[r]{23/2}}
\put(-17,45){\makebox(5,0)[r]{29/2}}
\put(-17,60){\makebox(5,0)[r]{35/2}}
\put(-17,75){\makebox(5,0)[r]{41/2}}

\put(50,-2.5){\separation}

\put(20,-7.5){\makebox(0,0){(a)}}

\put(0,20){\usebox{\four}}
\put(5,18){\makebox(0,0){0}}
\put(5,33){\makebox(0,0){3}}
\put(5,48){\makebox(0,0){6}}
\put(5,63){\makebox(0,0){9}}
\put(5,67){\usebox{\dash}}

\put(15,40){\usebox{\three}}
\put(20,38){\makebox(0,0){1}}
\put(20,53){\makebox(0,0){4}}
\put(20,68){\makebox(0,0){7}}
\put(20,72){\usebox{\dash}}

\put(30,0){\usebox{\six}}
\put(35,-2){\makebox(0,0){2}}
\put(35,13){\makebox(0,0){5}}
\put(35,28){\makebox(0,0){8}}
\put(35,43){\makebox(0,0){11}}
\put(35,58){\makebox(0,0){14}}
\put(35,73){\makebox(0,0){17}}
\put(35,77){\usebox{\dash}}


\put(80,-7.5){\makebox(0,0){(b)}}

\put(60,27.5){\usebox{\four}}
\put(65,25.5){\makebox(0,0){0}}
\put(65,40.5){\makebox(0,0){3}}
\put(65,55.5){\makebox(0,0){6}}
\put(65,70.5){\makebox(0,0){9}}
\put(65,74.5){\usebox{\dash}}

\put(75,47.5){\usebox{\three}}
\put(80,45.5){\makebox(0,0){1}}
\put(80,60.5){\makebox(0,0){4}}
\put(80,75.5){\makebox(0,0){7}}
\put(80,79.5){\usebox{\dash}}

\put(90,0){\usebox{\six}}
\put(95,-2){\makebox(0,0){2}}
\put(95,13){\makebox(0,0){5}}
\put(95,28){\makebox(0,0){8}}
\put(95,43){\makebox(0,0){11}}
\put(95,58){\makebox(0,0){14}}
\put(95,73){\makebox(0,0){17}}
\put(95,77){\usebox{\dash}}

\end{picture}

\caption{Examples of nondegenerate $H_0$~spectra belonging to class~(III):
(a) type (III.1.2.2) spectrum with $\alpha_0 = 18$, $\alpha_1 = -12$; (b) type
(III.2.2.2) spectrum with $\alpha_0 = 21$, $\alpha_1 = -15$.}

\end{figure}
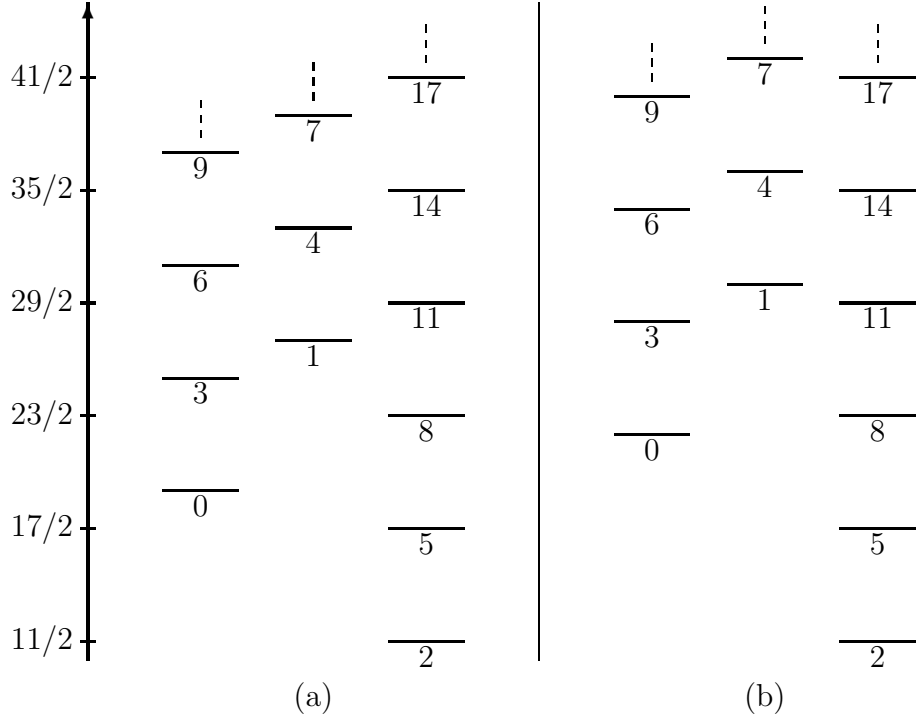
%
%
A similar procedure can be used for classes~(II) and~(III). Both of them
separate
into two subclasses (II.1), (II.2), and (III.1), (III.2), but the latter
are now labelled
by two integer indices $m$, $n=1$, 2, 3,~$\ldots$, instead of only one as for
class~(I). They are given by
\begin{eqnarray}
  & \mbox{(II.1.$m$.$n$)} & E_0 < E_3 < \cdots < E_{3n-3} < E_2 < E_{3n} < E_5 <
            \cdots < E_{3m+3n-6} \nonumber \\
  & & < E_{3m-1} < E_1 < E_{3m+3n-3} < E_{3m+2} < E_4 < \cdots \nonumber \\
  & & \mbox{if $6m - 4 < \alpha_0 < 6m + 2$ and $6n - 10 < \alpha_1 < 6m + 6n$}
            \nonumber \\
  & &  \mbox{} - \alpha_0 - 8, \label{eq:II-nondeg}  \\
  & \mbox{(II.2.$m$.$n$)} & E_0 < E_3 < \cdots < E_{3n-3} < E_2 < E_{3n} < E_5 <
            E_{3n+3} < \cdots \nonumber \\
  & & < E_{3m-1} < E_{3m+3n-3} < E_1 < E_{3m+2} < E_{3m+3n} < E_4 < \cdots
            \nonumber \\
  & & \mbox{if $6m - 4 < \alpha_0 < 6m + 2$ and $6m + 6n - \alpha_0 - 8 <
            \alpha_1$} \nonumber \\
  & &  < 6n - 4, \nonumber
\end{eqnarray}
and
\begin{eqnarray}
  & \mbox{(III.1.$m$.$n$)} & E_2 < E_5 < \cdots < E_{3n-1} < E_0 < E_{3n+2}
< E_3 <
            \cdots < E_{3m+3n-4} \nonumber \\
  & & < E_{3m-3} < E_1 < E_{3m+3n-1} < E_{3m} < E_4 < \cdots \nonumber \\
  & & \mbox{if $6m + 6n - 10 < \alpha_0 < 6m + 6n - 4$ and $6m - \alpha_0 - 8 <
            \alpha_1$} \nonumber \\
  & & < 2 - 6n, \label{eq:III-nondeg} \\
  & \mbox{(III.2.$m$.$n$)} & E_2 < E_5 < \cdots < E_{3n-1} < E_0 < E_{3n+2}
< E_3 <
            E_{3n+5} < \cdots  \nonumber \\
  & & < E_{3m-3} < E_{3m+3n-1} < E_1 < E_{3m} < E_{3m+3n+2} < E_4 < \cdots
            \nonumber \\
  & & \mbox{if $6m + 6n - 4 < \alpha_0 < 6m + 6n + 2$ and $- 4 - 6n < \alpha_1$}
            \nonumber \\
  & & < 6m - \alpha_0 - 2, \nonumber
\end{eqnarray}
respectively.\par
%
%
The parameter values given in equations~(\ref{eq:II-nondeg}),
and~(\ref{eq:III-nondeg}) can be checked in the same way as those in
equation~(\ref{eq:I-nondeg}). Furthermore, those corresponding to type
(II.1.$m$.$n$)
and (II.2.$m$.$n$) spectra cover all class~(II) parameter range, but for
$6m - 4 <
\alpha_0 < 6m + 2$, $\alpha_1 = 6n - 4$ or $\alpha_1 = 6m + 6n - \alpha_0 - 8$,
where $m$, $n=1$, 2, 3,~$\ldots$, and
$\alpha_1 > - 4$, $\alpha_0 = 6m + 2$, where $m=1$, 2, 3,~$\ldots$. A similar
remark applies to type (III.1.$m$.$n$) and (III.2.$m$.$n$) spectra, and
class~(III)
parameter range, the exceptions being now $6m + 6n - 4 < \alpha_0 < 6m + 6n
+ 2$,
$\alpha_1 = - 4 - 6n$ or $\alpha_1 = 6m - \alpha_0 - 2$, where $m$, $n=1$, 2,
3,~$\ldots$, and $- 2 - \alpha_0 < \alpha_1 < - 4$, $\alpha_0 = 6n + 2$, where
$n=1$, 2, 3,~$\ldots$.\par
%
%
Some examples of class (I), (II), and (III) nondegenerate spectra are
displayed on
figures~1, 2, and~3, respectively. One should remark that only type (I.1.1)
spectra,
for which $-1 < \alpha_0 < 2$ and $- 2 - \alpha_0 < \alpha_1 < 2$, have the same
level order as the standard harmonic oscillator, the spectrum of the latter being
retrieved in the special case where $\alpha_0 = \alpha_1 = 0$.\par
%
%
\subsection{Doubly-degenerate spectra}   \label{sec:doubly}
\hspace{\parindent}
Doubly-degenerate spectra may appear as limiting cases of the nondegenerate ones
of subsection~\ref{sec:nondeg}, whenever two contiguous energies become
equal, or
they may directly result from the two intermediate classes, defined in
equation~(\ref{eq:interm}). They belong to three different types, labelled
by a, b,~c,
and corresponding to ${\cal F}_0$--${\cal F}_1$, ${\cal F}_0$--${\cal F}_2$, and
${\cal F}_1$--${\cal F}_2$ degeneracies, respectively.\par
%
%
\begin{figure}

\begin{picture}(130,95)(-37,-10)

\put(-10,-2.5){\thicklines\vector(0,1){87.5}}
\multiput(-11,0)(0,15){6}{\ord}
\put(-17,0){\makebox(5,0)[r]{1/2}}
\put(-17,15){\makebox(5,0)[r]{7/2}}
\put(-17,30){\makebox(5,0)[r]{13/2}}
\put(-17,45){\makebox(5,0)[r]{19/2}}
\put(-17,60){\makebox(5,0)[r]{25/2}}
\put(-17,75){\makebox(5,0)[r]{31/2}}

\put(50,-2.5){\separation}

\put(20,-7.5){\makebox(0,0){(a)}}

\put(0,0){\usebox{\six}}
\put(5,-2){\makebox(0,0){0}}
\put(5,13){\makebox(0,0){3}}
\put(5,28){\makebox(0,0){6}}
\put(5,43){\makebox(0,0){9}}
\put(5,58){\makebox(0,0){12}}
\put(5,73){\makebox(0,0){15}}
\put(5,77){\usebox{\dash}}

\put(15,30){\usebox{\four}}
\put(20,28){\makebox(0,0){1}}
\put(20,43){\makebox(0,0){4}}
\put(20,58){\makebox(0,0){7}}
\put(20,73){\makebox(0,0){10}}
\put(20,77){\usebox{\dash}}

\put(30,35){\usebox{\three}}
\put(35,33){\makebox(0,0){2}}
\put(35,48){\makebox(0,0){5}}
\put(35,63){\makebox(0,0){8}}
\put(35,67){\usebox{\dash}}


\put(80,-7.5){\makebox(0,0){(b)}}

\put(60,0){\usebox{\six}}
\put(65,-2){\makebox(0,0){0}}
\put(65,13){\makebox(0,0){3}}
\put(65,28){\makebox(0,0){6}}
\put(65,43){\makebox(0,0){ 9}}
\put(65,58){\makebox(0,0){12}}
\put(65,73){\makebox(0,0){15}}
\put(65,77){\usebox{\dash}}

\put(75,25){\usebox{\four}}
\put(80,23){\makebox(0,0){1}}
\put(80,38){\makebox(0,0){4}}
\put(80,53){\makebox(0,0){7}}
\put(80,68){\makebox(0,0){10}}
\put(80,72){\usebox{\dash}}

\put(90,30){\usebox{\four}}
\put(95,28){\makebox(0,0){2}}
\put(95,43){\makebox(0,0){5}}
\put(95,58){\makebox(0,0){8}}
\put(95,73){\makebox(0,0){11}}
\put(95,77){\usebox{\dash}}

\end{picture}

\caption{Examples of doubly-degenerate $H_0$~spectra belonging to class~(I):
(a) type (I.2.a) spectrum with $\alpha_0 = 0$, $\alpha_1 = 10$; (b) type (I.2.b)
spectrum with $\alpha_0 = 0$, $\alpha_1 = 8$.}

\end{figure}
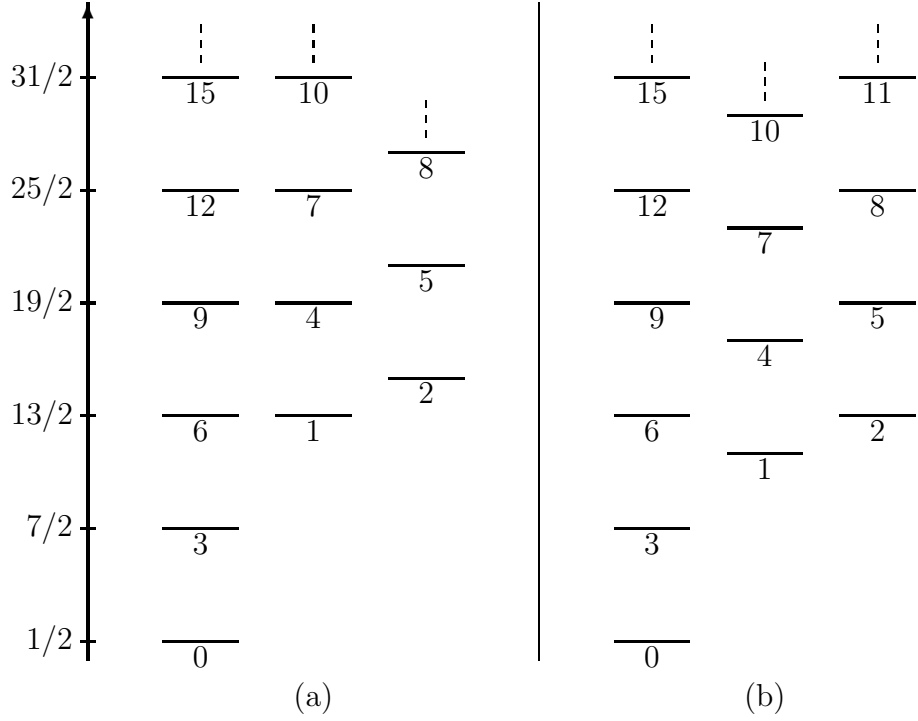
%
%
{}For class (I), for instance, we can obtain type~a spectra by considering
the limit
$E_1 = E_{3n}$ in subclass~(I.2.$n$), defined in equation~(\ref{eq:I-nondeg}),
thereby getting the condition $\alpha_1 = 6n - \alpha_0 - 2$. The remaining two
possibilities, namely $E_{3n-3} = E_1$ in subclass~(I.1.$n$) or~(I.2.$n$)
for $n=2$,
3,~$\ldots$, can be excluded because the former leads to the same types of
spectra and parameter values as those already found, while the latter would
imply
the $\alpha_1$ value $6n - \alpha_0 - 8$, lying outside the interval $(6n -
4, 6n -
\alpha_0 - 2)$. Similarly, type~b spectra can be obtained by considering
the limit
$E_2 = E_{3n}$ in subclass~(I.1.$n$) or~(I.2.$n$), thus giving the
condition $\alpha_1
= 6n - 4$. On the contrary, type~c spectra cannot be derived as limiting
cases of
class~(I) spectra, as $E_1 < E_2$ by definition of the class, and $E_2 <
E_4$ over
the whole parameter range.\par
%
%
\renewcommand{\separation}{\line(0,1){92.5}}

\begin{figure}

\begin{picture}(130,100)(-18,-15)

\put(-5,-7.5){\thicklines\vector(0,1){92.5}}
\multiput(-6,0)(0,15){6}{\ord}
\put(-12,0){\makebox(5,0)[r]{11/2}}
\put(-12,15){\makebox(5,0)[r]{17/2}}
\put(-12,30){\makebox(5,0)[r]{23/2}}
\put(-12,45){\makebox(5,0)[r]{29/2}}
\put(-12,60){\makebox(5,0)[r]{35/2}}
\put(-12,75){\makebox(5,0)[r]{41/2}}

\multiput(45,-7.5)(50,0){2}{\separation}

\put(20,-12.5){\makebox(0,0){(a)}}

\put(0,0){\usebox{\six}}
\put(5,-2){\makebox(0,0){0}}
\put(5,13){\makebox(0,0){3}}
\put(5,28){\makebox(0,0){6}}
\put(5,43){\makebox(0,0){9}}
\put(5,58){\makebox(0,0){12}}
\put(5,73){\makebox(0,0){15}}
\put(5,77){\usebox{\dash}}

\put(15,45){\usebox{\three}}
\put(20,43){\makebox(0,0){1}}
\put(20,58){\makebox(0,0){4}}
\put(20,73){\makebox(0,0){7}}
\put(20,77){\usebox{\dash}}

\put(30,25){\usebox{\four}}
\put(35,23){\makebox(0,0){2}}
\put(35,38){\makebox(0,0){5}}
\put(35,53){\makebox(0,0){8}}
\put(35,68){\makebox(0,0){11}}
\put(35,72){\usebox{\dash}}


\put(70,-12.5){\makebox(0,0){(b)}}

\put(50,0){\usebox{\six}}
\put(55,-2){\makebox(0,0){0}}
\put(55,13){\makebox(0,0){3}}
\put(55,28){\makebox(0,0){6}}
\put(55,43){\makebox(0,0){9}}
\put(55,58){\makebox(0,0){12}}
\put(55,73){\makebox(0,0){15}}
\put(55,77){\usebox{\dash}}

\put(65,35){\usebox{\three}}
\put(70,33){\makebox(0,0){1}}
\put(70,48){\makebox(0,0){4}}
\put(70,63){\makebox(0,0){7}}
\put(70,67){\usebox{\dash}}

\put(80,15){\usebox{\five}}
\put(85,13){\makebox(0,0){2}}
\put(85,28){\makebox(0,0){5}}
\put(85,43){\makebox(0,0){8}}
\put(85,58){\makebox(0,0){11}}
\put(85,73){\makebox(0,0){14}}
\put(85,77){\usebox{\dash}}


\put(120,-12.5){\makebox(0,0){(c)}}

\put(100,-5){\usebox{\six}}
\put(105,-7){\makebox(0,0){0}}
\put(105,8){\makebox(0,0){3}}
\put(105,23){\makebox(0,0){6}}
\put(105,38){\makebox(0,0){9}}
\put(105,53){\makebox(0,0){12}}
\put(105,68){\makebox(0,0){15}}
\put(105,72){\usebox{\dash}}

\put(115,30){\usebox{\four}}
\put(120,28){\makebox(0,0){1}}
\put(120,43){\makebox(0,0){4}}
\put(120,58){\makebox(0,0){7}}
\put(120,73){\makebox(0,0){10}}
\put(120,77){\usebox{\dash}}

\put(130,15){\usebox{\five}}
\put(135,13){\makebox(0,0){2}}
\put(135,28){\makebox(0,0){5}}
\put(135,43){\makebox(0,0){8}}
\put(135,58){\makebox(0,0){11}}
\put(135,73){\makebox(0,0){14}}
\put(135,77){\usebox{\dash}}

\end{picture}

\caption{Examples of doubly-degenerate $H_0$~spectra belonging to
class~(II): (a) type (II.2.2.a) spectrum with $\alpha_0 = 10$, $\alpha_1 =
6$; (b)
type (II.2.2.b) spectrum with $\alpha_0 = 10$, $\alpha_1 = 2$; (c) type
(II.2.2.c)
spectrum with $\alpha_0 = 8$, $\alpha_1 = 4$.}

\end{figure}
%
%
%
\renewcommand{\separation}{\line(0,1){92.5}}

\begin{figure}

\begin{picture}(130,100)(-18,-15)

\put(-5,-7.5){\thicklines\vector(0,1){92.5}}
\multiput(-6,0)(0,15){6}{\ord}
\put(-12,0){\makebox(5,0)[r]{15/2}}
\put(-12,15){\makebox(5,0)[r]{21/2}}
\put(-12,30){\makebox(5,0)[r]{27/2}}
\put(-12,45){\makebox(5,0)[r]{33/2}}
\put(-12,60){\makebox(5,0)[r]{39/2}}
\put(-12,75){\makebox(5,0)[r]{45/2}}

\multiput(45,-7.5)(50,0){2}{\separation}

\put(20,-12.5){\makebox(0,0){(a)}}

\put(0,25){\usebox{\four}}
\put(5,23){\makebox(0,0){0}}
\put(5,38){\makebox(0,0){3}}
\put(5,53){\makebox(0,0){6}}
\put(5,68){\makebox(0,0){9}}
\put(5,72){\usebox{\dash}}

\put(15,55){\usebox{\two}}
\put(20,53){\makebox(0,0){1}}
\put(20,68){\makebox(0,0){4}}
\put(20,72){\usebox{\dash}}

\put(30,0){\usebox{\six}}
\put(35,-2){\makebox(0,0){2}}
\put(35,13){\makebox(0,0){5}}
\put(35,28){\makebox(0,0){ 8}}
\put(35,43){\makebox(0,0){11}}
\put(35,58){\makebox(0,0){14}}
\put(35,73){\makebox(0,0){17}}
\put(35,77){\usebox{\dash}}


\put(70,-12.5){\makebox(0,0){(b)}}

\put(50,25){\usebox{\four}}
\put(55,23){\makebox(0,0){0}}
\put(55,38){\makebox(0,0){3}}
\put(55,53){\makebox(0,0){6}}
\put(55,68){\makebox(0,0){9}}
\put(55,72){\usebox{\dash}}

\put(65,50){\usebox{\two}}
\put(70,48){\makebox(0,0){1}}
\put(70,63){\makebox(0,0){4}}
\put(70,67){\usebox{\dash}}

\put(80,-5){\usebox{\six}}
\put(85,-7){\makebox(0,0){2}}
\put(85,8){\makebox(0,0){5}}
\put(85,23){\makebox(0,0){8}}
\put(85,38){\makebox(0,0){11}}
\put(85,53){\makebox(0,0){14}}
\put(85,68){\makebox(0,0){17}}
\put(85,72){\usebox{\dash}}


\put(120,-12.5){\makebox(0,0){(c)}}

\put(100,15){\usebox{\five}}
\put(105,13){\makebox(0,0){0}}
\put(105,28){\makebox(0,0){3}}
\put(105,43){\makebox(0,0){6}}
\put(105,58){\makebox(0,0){9}}
\put(105,73){\makebox(0,0){12}}
\put(105,77){\usebox{\dash}}

\put(115,40){\usebox{\three}}
\put(120,38){\makebox(0,0){1}}
\put(120,53){\makebox(0,0){4}}
\put(120,68){\makebox(0,0){7}}
\put(120,72){\usebox{\dash}}

\put(130,-5){\usebox{\six}}
\put(135,-7){\makebox(0,0){2}}
\put(135,8){\makebox(0,0){5}}
\put(135,23){\makebox(0,0){8}}
\put(135,38){\makebox(0,0){11}}
\put(135,53){\makebox(0,0){14}}
\put(135,68){\makebox(0,0){17}}
\put(135,72){\usebox{\dash}}

\end{picture}

\caption{Examples of doubly-degenerate $H_0$~spectra belonging to
class~(III): (a) type (III.2.2.a) spectrum with $\alpha_0 = 24$, $\alpha_1
= -14$; (b)
type (III.2.2.b) spectrum with $\alpha_0 = 24$, $\alpha_1 = -16$; (c) type
(III.2.2.c)
spectrum with $\alpha_0 = 20$, $\alpha_1 = -12$.}

\end{figure}
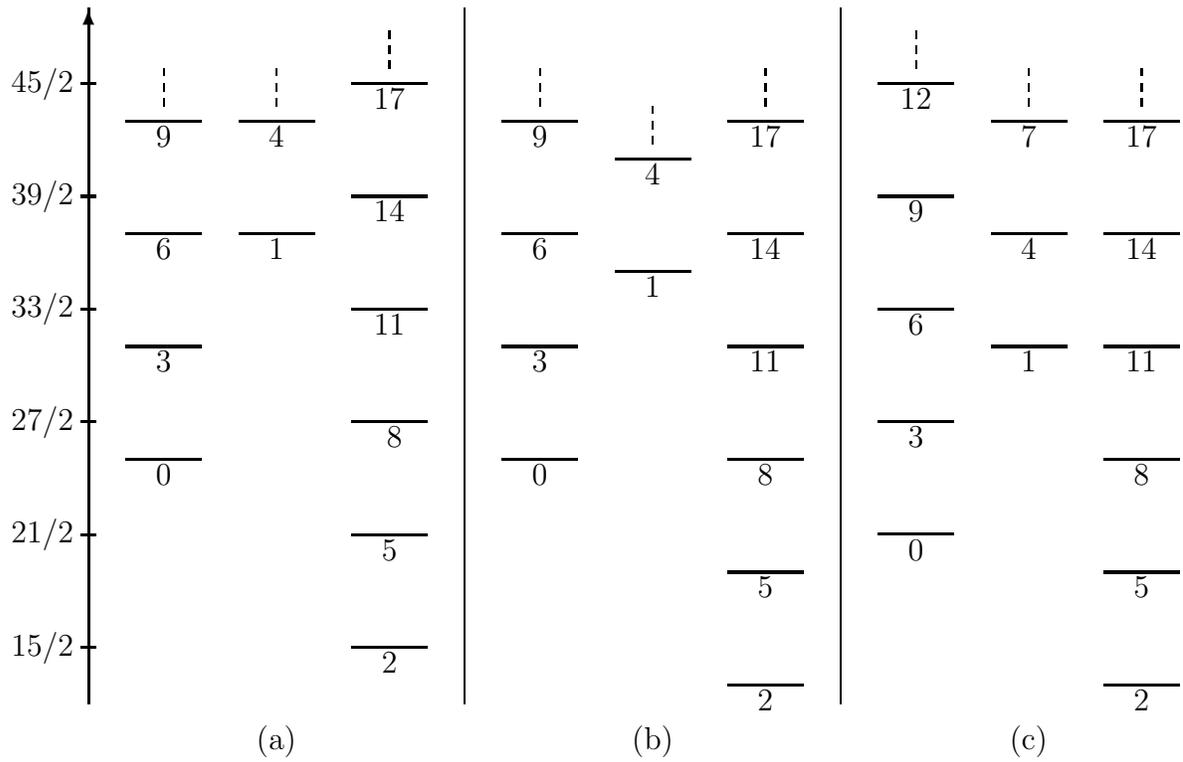
%
%
We conclude that class (I) doubly-degenerate spectra are given by
\begin{eqnarray}
  & \mbox{(I.$n$.a)} & E_0 < E_3 < \cdots < E_{3n-3} < E_{3n} = E_1 < E_2 <
E_{3n+3} =
            E_4 \nonumber \\
  & & < E_5 < \cdots \nonumber \\
  & & \mbox{if $-1 < \alpha_0 < 2$ and $\alpha_1 = 6n - \alpha_0 - 2$},
\nonumber
            \\
  & \mbox{(I.$n$.b)} & E_0 < E_3 < \cdots < E_{3n-3} < E_1 < E_{3n} = E_2 <
E_4 \\
  & & < E_{3n+3} = E_5 < \cdots \nonumber \\
  & & \mbox{if $-1 < \alpha_0 < 2$ and $\alpha_1 = 6n - 4$}, \nonumber
\end{eqnarray}
where $n$ runs over 1, 2, 3,~$\ldots$. Together with type (I.1.$n$) and
(I.2.$n$)
nondegenerate spectra, they clearly exhaust all class~(I) spectra.\par
%
%
By proceeding in the same way, the doubly-degenerate spectra, arising as
limiting
cases of class~(III) nondegenerate ones, can be shown to separate into the
following types:
\begin{eqnarray}
    & \mbox{(III.$m$.$n$.a)} & E_2 < E_5 < \cdots < E_{3n-1} < E_0 <
E_{3n+2} < E_3 <
            E_{3n+5} < \cdots \nonumber \\
    & & < E_{3m-3} < E_{3m+3n-1} < E_{3m} = E_1 < E_{3m+3n+2} \nonumber \\
    & & < E_{3m+3} = E_4 < \cdots \nonumber \\
    & & \mbox{if $6m + 6n - 4 < \alpha_0 < 6m + 6n +2$ and $\alpha_1 = 6m -
            \alpha_0 - 2$}, \nonumber \\
    & \mbox{(III.$m$.$n$.b)} & E_2 < E_5 < \cdots < E_{3n-1} < E_{3n+2} = E_0 <
            E_{3n+5} = E_3 < \cdots \nonumber \\
    & & < E_{3m+3n-1} = E_{3m-3} < E_1 < E_{3m+3n+2} = E_{3m} < E_4 < \cdots \\
    & & \mbox{if $6m + 6n - 4 < \alpha_0 < 6m + 6n +2$ and $\alpha_1 =  - 4
- 6n$},
            \nonumber \\
    & \mbox{(III.$m$.$n$.c)} & E_2 < E_5 < \cdots < E_{3n-1} < E_0 <
E_{3n+2} < E_3 <
             \cdots \nonumber \\
    & & < E_{3m+3n-4} < E_{3m-3} < E_{3m+3n-1} = E_1 < E_{3m} \nonumber \\
    & & < E_{3m+3n+2} = E_4 < \cdots \nonumber \\
    & & \mbox{if $\alpha_0 = 6m + 6n - 4$ and $- 4 - 6n < \alpha_1 < 2 - 6n$},
             \nonumber
\end{eqnarray}
where $m$, $n$ run over 1, 2, 3,~$\ldots$. Together with
type~(III.1.$m$.$n$) and
(III.2.$m$.$n$) nondegenerate spectra, they cover all class~(III) parameter
range,
but for the discrete values $\alpha_0 = 6m + 6n + 2$, $\alpha_1 = - 4 -
6n$, where
$m$, $n=1$, 2, 3,~$\ldots$.\par
%
%
The doubly-degenerate spectra arising as limiting cases of class~(II)
nondegenerate ones can be grouped with those appearing in the intermediate
classes (I-II), and (II-III) to provide the following types:
\begin{eqnarray}
    & \mbox{(II.$m$.$n$.a)} & E_0 < E_3 < \cdots < E_{3n-3} < E_2 < E_{3n}
< E_5 <
            \cdots < E_{3m+3n-6}  \nonumber \\
    & & < E_{3m-1} < E_{3m+3n-3} = E_1 < E_{3m+2} < E_{3m+3n} = E_4 <
            \cdots \nonumber \\
    & & \mbox{if $6m - 4 < \alpha_0 < 6m +2$ and $\alpha_1 = 6m + 6n - \alpha_0
            - 8$}, \nonumber \\
    & \mbox{(II.$m$.$n$.b)} & E_0 < E_3 < \cdots < E_{3n-3} = E_2 < E_{3n}
= E_5 <
            \cdots \nonumber \\
    & &< E_{3m+3n-6} = E_{3m-1} < E_1 < E_{3m+3n-3} = E_{3m+2} < E_4 < \cdots
            \nonumber \\
    & & \mbox{if $6m - 4 < \alpha_0 < 6m +2$ and $\alpha_1 =  6n - 10$}, \\
    & \mbox{(II.$m$.$n$.c)} & E_0 < E_3 < \cdots < E_{3n-3} < E_2 < E_{3n}
< E_5 <
             E_{3n+3} <\cdots \nonumber \\
    & & < E_{3m-4} < E_{3m+3n-6} < E_{3m-1} = E_1 < E_{3m+3n-3} \nonumber \\
    & & < E_{3m+2} = E_4 < \cdots \nonumber \\
    & & \mbox{if $\alpha_0 = 6m - 4$ and $6n - 10 < \alpha_1 < 6n - 4$},
             \nonumber
\end{eqnarray}
where $m$, $n$ run over 1, 2, 3,~$\ldots$. Here we note that type
(II.$m$.$n$.a),
(II.$m$.$n$.b) (with $n\ge 2$), and (II.$m$.$n$.c) (with $m\ge 2$) spectra
come from
class~(II), and together with type (II.1.$m$.$n$) and (II.2.$m$.$n$)
nondegenerate
spectra cover all class~(II) parameter range, but for the discrete values
$\alpha_0 =
6m + 2$, $\alpha_1 = 6n - 4$, where $m$, $n=1$, 2, 3,~$\ldots$. On the contrary,
type (II.1.$n$.c) [resp.~(II.$m$.1.b)] spectra result from the intermediate
class~(I-II)
[resp.~(II-III)], and cover all the corresponding parameter range, but for the
discrete values $\alpha_0 = 2$, $\alpha_1 = 6n - 4$, where $n=1$, 2, 3,~$\ldots$
[resp.~$\alpha_1 = -4$, $\alpha_0 = 6m + 2$, where $m=1$, 2, 3,~$\ldots$].\par
%
%
Some examples of doubly-degenerate spectra are displayed on figures~4, 5, and~6.
One should note that the lowest doubly-degenerate state is the $k$th one,
where $k
= n+1$, $n+2$, $2m+n$, $n$, $2m+n-1$, $2m+n+1$, $n+1$, or $2m+n$ for type
(I.$n$.a), (I.$n$.b), (II.$m$.$n$.a), (II.$m$.$n$.b), (II.$m$.$n$.c),
(III.$m$.$n$.a),
(III.$m$.$n$.b), or (III.$m$.$n$.c), respectively, and that above such a
doubly-degenerate state, there always remain some nondegenerate ones. For
type~(II.$m$.1.b) spectra, and only for them, the ground state is doubly
degenerate.\par
%
%
\subsection{Triply-degenerate spectra}   \label{sec:triply}
\hspace{\parindent}
The allowed parameter values not encountered in subsections~\ref{sec:nondeg},
\ref{sec:doubly} correspond to triply-degenerate spectra. The latter may be
separated into the following three types:
\begin{eqnarray}
    & \mbox{(I.$n$.abc)} & E_0 < E_3 < \cdots < E_{3n-3} < E_{3n} = E_1 = E_2
            \nonumber \\
    & & < E_{3n+3} = E_4 = E_5 < \cdots \nonumber \\
    & & \mbox{if $\alpha_0 = 2$ and $\alpha_1 = 6n - 4$}, \nonumber \\
    & \mbox{(II.$m$.$n$.abc)} & E_0 < E_3 < \cdots < E_{3n-6} < E_{3n-3} = E_2 <
            E_{3n} = E_5 < \cdots \nonumber \\
    & & < E_{3m+3n-6} = E_{3m-1} < E_{3m+3n-3} = E_{3m+2} = E_1 \nonumber \\
    & & < E_{3m+3n} = E_{3m+5} =  E_4 < \cdots \\
    & & \mbox{if $\alpha_0 = 6m +2$ and $\alpha_1 =  6n - 10$}, \nonumber \\
    & \mbox{(III.$m$.$n$.abc)} & E_2 < E_5 < \cdots < E_{3n-1} < E_{3n+2} =
E_0 <
            E_{3n+5} = E_3 < \cdots \nonumber \\
    & & < E_{3m+3n-1} = E_{3m-3} < E_{3m+3n+2} = E_{3m} = E_1  \nonumber \\
    & & < E_{3m+3n+5} = E_{3m+3} = E_4 < \cdots \nonumber \\
    & & \mbox{if $\alpha_0 = 6m + 6n +2$ and $\alpha_1 = - 4 - 6n$}, \nonumber
\end{eqnarray}
where $m$, $n$ run over 1, 2, 3,~$\ldots$. The first type comes from the
intermediate class (I-II), the second one from class~(II) or from the
intermediate
class~(II-III), according to whether $n \ge 2$ or $n=1$, while the third
one results
from class~(III).\par
%
%
\renewcommand{\separation}{\line(0,1){87.5}}

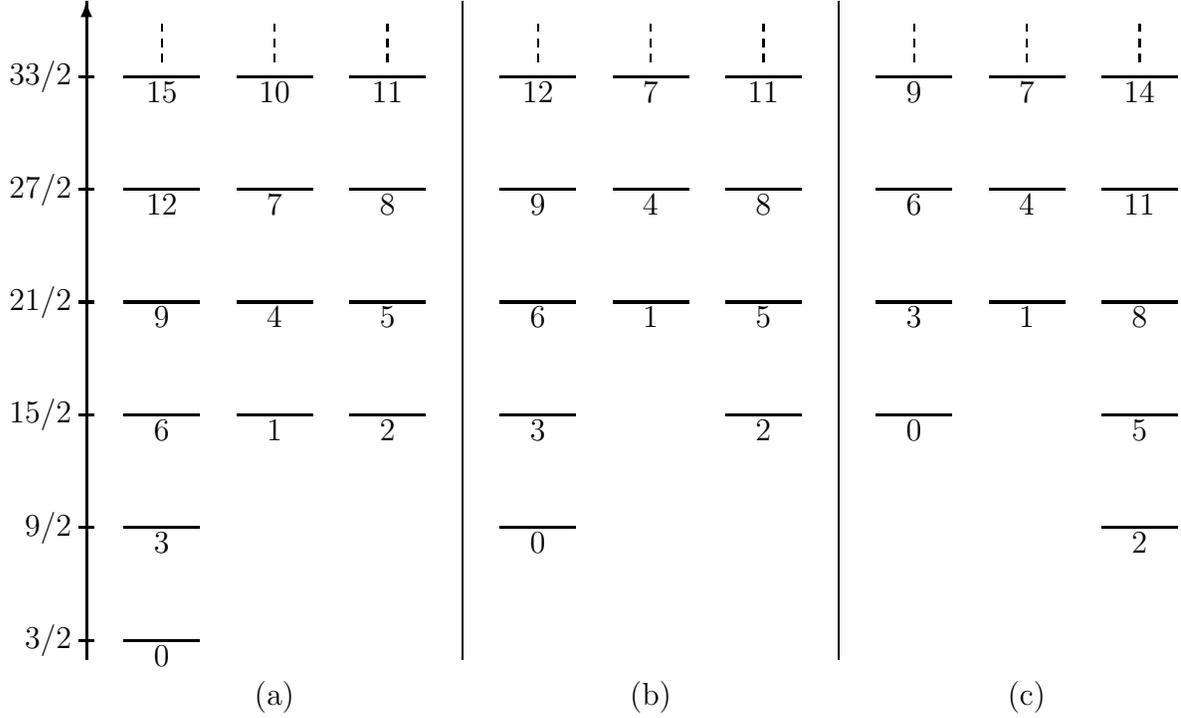
\begin{figure}

\begin{picture}(130,95)(-18,-10)

\put(-5,-2.5){\thicklines\vector(0,1){87.5}}
\multiput(-6,0)(0,15){6}{\ord}
\put(-12,0){\makebox(5,0)[r]{3/2}}
\put(-12,15){\makebox(5,0)[r]{9/2}}
\put(-12,30){\makebox(5,0)[r]{15/2}}
\put(-12,45){\makebox(5,0)[r]{21/2}}
\put(-12,60){\makebox(5,0)[r]{27/2}}
\put(-12,75){\makebox(5,0)[r]{33/2}}

\multiput(45,-2.5)(50,0){2}{\separation}

\put(20,-7.5){\makebox(0,0){(a)}}

\put(0,0){\usebox{\six}}
\put(5,-2){\makebox(0,0){0}}
\put(5,13){\makebox(0,0){3}}
\put(5,28){\makebox(0,0){6}}
\put(5,43){\makebox(0,0){9}}
\put(5,58){\makebox(0,0){12}}
\put(5,73){\makebox(0,0){15}}
\put(5,77){\usebox{\dash}}

\put(15,30){\usebox{\four}}
\put(20,28){\makebox(0,0){1}}
\put(20,43){\makebox(0,0){4}}
\put(20,58){\makebox(0,0){7}}
\put(20,73){\makebox(0,0){10}}
\put(20,77){\usebox{\dash}}

\put(30,30){\usebox{\four}}
\put(35,28){\makebox(0,0){2}}
\put(35,43){\makebox(0,0){5}}
\put(35,58){\makebox(0,0){8}}
\put(35,73){\makebox(0,0){11}}
\put(35,77){\usebox{\dash}}


\put(70,-7.5){\makebox(0,0){(b)}}

\put(50,15){\usebox{\five}}
\put(55,13){\makebox(0,0){0}}
\put(55,28){\makebox(0,0){3}}
\put(55,43){\makebox(0,0){6}}
\put(55,58){\makebox(0,0){9}}
\put(55,73){\makebox(0,0){12}}
\put(55,77){\usebox{\dash}}

\put(65,45){\usebox{\three}}
\put(70,43){\makebox(0,0){1}}
\put(70,58){\makebox(0,0){4}}
\put(70,73){\makebox(0,0){7}}
\put(70,77){\usebox{\dash}}

\put(80,30){\usebox{\four}}
\put(85,28){\makebox(0,0){2}}
\put(85,43){\makebox(0,0){5}}
\put(85,58){\makebox(0,0){8}}
\put(85,73){\makebox(0,0){11}}
\put(85,77){\usebox{\dash}}


\put(120,-7.5){\makebox(0,0){(c)}}

\put(100,30){\usebox{\four}}
\put(105,28){\makebox(0,0){0}}
\put(105,43){\makebox(0,0){3}}
\put(105,58){\makebox(0,0){6}}
\put(105,73){\makebox(0,0){9}}
\put(105,77){\usebox{\dash}}

\put(115,45){\usebox{\three}}
\put(120,43){\makebox(0,0){1}}
\put(120,58){\makebox(0,0){4}}
\put(120,73){\makebox(0,0){7}}
\put(120,77){\usebox{\dash}}

\put(130,15){\usebox{\five}}
\put(135,13){\makebox(0,0){2}}
\put(135,28){\makebox(0,0){5}}
\put(135,43){\makebox(0,0){8}}
\put(135,58){\makebox(0,0){11}}
\put(135,73){\makebox(0,0){14}}
\put(135,77){\usebox{\dash}}

\end{picture}

\caption{Examples of triply-degenerate $H_0$~spectra: (a) type
(I.2.abc) spectrum with $\alpha_0 = 2$, $\alpha_1 = 8$; (b) type (II.1.2.abc)
spectrum with $\alpha_0 = 8$, $\alpha_1 = 2$; (c) type (III.1.1.abc)
spectrum with
$\alpha_0 = 14$, $\alpha_1 = -10$.}

\end{figure}
%
%
Some examples of triply-degenerate spectra are displayed on figure~7. Below the
infinite set of triply-degenerate states, there appear $n$ nondegenerate
states in
type~(I.$n$.abc) spectra, while in the case of type~(II.$m$.$n$.abc)
[resp.~(III.$m$.$n$.abc)] spectra, there are $n-1$ [resp.~$n$]
nondegenerate states,
followed by $m$ [resp.~$m$] doubly-degenerate ones. For type~(II.$m$.1.abc)
spectra, and only for them, the ground state is doubly degenerate. No
spectrum with
a triply-degenerate ground state is obtained.\par
%
%
\section{Period-Three Spectra and Supersymmetric Quantum Mechanics}
\label{sec:periodic}
\setcounter{equation}{0}
\hspace{\parindent}
{}From equations~(\ref{eq:I-nondeg}), (\ref{eq:II-nondeg}),
and~(\ref{eq:III-nondeg}),
it results that type (I.1.1), (II.1.1.1), and (III.1.1.1) spectra,
characterized by
\begin{eqnarray}
  & \mbox{(I.1.1)} & E_0 < E_1 < E_2 < E_3 < E_4 < E_5 < \cdots \nonumber \\
  & & \mbox{if $-1 < \alpha_0 < 2$ and $- 2 - \alpha_0 < \alpha_1 < 2$},
            \nonumber \\
  & \mbox{(II.1.1.1)} & E_0 < E_2 < E_1 < E_3 < E_5 < E_4 < \cdots \nonumber \\
  & & \mbox{if $2 < \alpha_0 < 8$ and $- 4 < \alpha_1 < 4 - \alpha_0$},
             \label{eq:period-spec}\\
  & \mbox{(III.1.1.1)} & E_2 < E_0 < E_1 < E_5 < E_3 < E_4 < \cdots \nonumber \\
  & & \mbox{if $2 < \alpha_0 < 8$ and $- 2 - \alpha_0 < \alpha_1 < - 4$},
             \nonumber
\end{eqnarray}
respectively, have an infinite number of periodically spaced levels. More
precisely,
the level spacings are given by $\omega_0$, $\omega_1$, $\omega_2$,
$\omega_0$, $\omega_1$, $\omega_2$,~$\ldots$, where $\omega_{\mu}$,
$\mu=0$, 1,~2, can be expressed in terms of the algebra parameters $\alpha_0$,
$\alpha_1$, as
\begin{eqnarray}
  & \mbox{(I.1.1)} & \omega_0 = \half (\alpha_0 + \alpha_1 + 2), \quad
\omega_1 =
          \half (2 - \alpha_0), \quad \omega_2 = \half (2 - \alpha_1),
\nonumber \\
  & \mbox{(II.1.1.1)} & \omega_0 = \half (\alpha_1 + 4), \quad \omega_1 =
          \half (\alpha_0 - 2), \quad \omega_2 = \half (4 - \alpha_0 -
\alpha_1), \\
  & \mbox{(III.1.1.1)} & \omega_0 = \half (- \alpha_1 - 4), \quad \omega_1 =
          \half (\alpha_0 + \alpha_1 + 2), \quad \omega_2 = \half (8 -
\alpha_0),
          \nonumber
\end{eqnarray}
respectively. In all three cases, the normalization of~$H_0$ is such that
$\Omega_3 \equiv \omega_0 + \omega_1 + \omega_2 = 3$.\par
%
%
Spectra of a similar type were recently encountered by Sukhatme {\sl et
al\/}~\cite{sukhatme} in the context of SSQM with cyclic shape invariant
potentials
of period three. In such a case, one may construct a hierarchy of supersymmetric
Hamiltonians, and corresponding supercharges in terms of superpotentials that
repeat after a cycle of three iterations. In terms of the operators
\begin{equation}
  A_{\mu} = \frac{d}{dx} + W(x,b_{\mu}), \qquad \Ap_{\mu} = - \frac{d}{dx} +
  W(x,b_{\mu}), \qquad \mu = 0, 1, 2, \ldots,
\end{equation}
where $b_{\mu}$ denotes a set of parameters such that $b_{\mu+3} = b_{\mu}$, and
the superpotentials $W(x,b_{\mu})$ satisfy the shape invariance conditions
\begin{equation}
  W^2(x,b_{\mu}) + W'(x,b_{\mu}) = W^2(x,b_{\mu+1}) - W'(x,b_{\mu+1}) +
  \omega_{\mu}, \qquad \mu = 0, 1, 2,
\end{equation}
the supersymmetric Hamiltonians~${\cal H}_{\mu}$, and supercharge operators
$\Qp_{\mu}$, $Q_{\mu}$ are defined by
\begin{equation}
  {\cal H}_{\mu} = \left(\begin{array}{cc}
                 {\cal H}^{(\mu)} - {\cal E}^{(\mu)}_0 I & 0 \\
                 0 & {\cal H}^{(\mu+1)} - {\cal E}^{(\mu)}_0 I
                            \end{array}\right), \quad
  \Qp_{\mu} = \left(\begin{array}{cc}
                 0 & \Ap_{\mu} \\
                 0 & 0
                     \end{array}\right), \quad
  Q_{\mu} = \left(\begin{array}{cc}
                 0 & 0 \\
                 A_{\mu} & 0
                  \end{array}\right),     \label{eq:super-op}
\end{equation}
where
\begin{eqnarray}
  {\cal H}^{(0)} & = & \Ap_0 A_0, \nonumber \\
  {\cal H}^{(\mu)} & = & A_{\mu-1} \Ap_{\mu-1} + {\cal E}^{(\mu-1)}_0 I =
\Ap_{\mu}
          A_{\mu} + {\cal E}^{(\mu)}_0 I, \qquad \mu = 1, 2, \ldots,
          \label{eq:hierarchy}
\end{eqnarray}
and ${\cal E}^{(\mu)}_0$ denotes the ground state energy of~${\cal H}^{(\mu)}$
(with ${\cal E}^{(0)}_0 = 0$).\par
%
%
Since $\Ap_3 = \Ap_0$, $A_3 = A_0$, ${\cal H}^{(3)} = {\cal H}^{(0)} + {\cal
E}^{(3)}_0 I$, one finds
\begin{equation}
  {\cal H}_{\mu+3} = {\cal H}_{\mu}, \qquad \Qp_{\mu+3} = \Qp_{\mu}, \qquad
  Q_{\mu+3} = Q_{\mu}.   \label{eq:period-op}
\end{equation}
Hence, there are only three sets of independent operators $\{ {\cal H}_{\mu},
\Qp_{\mu}, Q_{\mu} \}$, corresponding to $\mu=0$, 1,~2. Each one of them fulfils
the defining relations of the sqm(2) superalgebra
\begin{equation}
  \left(\Qp_{\mu}\right)^2 = Q_{\mu}^2 = 0, \qquad \left[{\cal H}_{\mu},
\Qp_{\mu}
  \right] = \left[{\cal H}_{\mu}, Q_{\mu}\right] = 0, \qquad \left\{Q_{\mu},
  \Qp_{\mu}\right\} = {\cal H}_{\mu}.    \label{eq:sqm}
\end{equation}
The eigenvalues ${\cal E}^{(\mu)}_n$, $n=0$, 1, 2,~$\ldots$, of ${\cal
H}^{(\mu)}$,
$\mu=0$, 1,~2, satisfy the relations
\begin{eqnarray}
  {\cal E}^{(0)}_1 & = & {\cal E}^{(1)}_0 = \omega_0, \nonumber \\
  {\cal E}^{(0)}_2 & = & {\cal E}^{(1)}_1 = {\cal E}^{(2)}_0 = \omega_0 +
\omega_1,
            \label{eq:super-spec} \\
  {\cal E}^{(0)}_{3k+\nu} & = & {\cal E}^{(1)}_{3k+\nu-1} = {\cal
E}^{(2)}_{3k+\nu-2} =
            {\cal E}^{(3)}_{3(k-1)+\nu} = k \Omega_3 + \sum_{\rho=0}^{\nu-1}
            \omega_{\rho}, \nonumber
\end{eqnarray}
where $k=1$, 2,~$\ldots$, $\nu=0$, 1,~2, and $\sum_{\rho=0}^{-1} \equiv 0$.\par
%
%
We shall now proceed to show that one may realize the operators defined in
equations~(\ref{eq:super-op}), (\ref{eq:hierarchy}), and satisfying
equations~(\ref{eq:period-op}), (\ref{eq:sqm}), in terms of creation and
annihilation
operators $\ap_{\mu}$, $a_{\mu}$, $\mu=0$, 1,~2, belonging to $C_3$-extended
oscillator algebras ${\cal A}^{(3)}_{\alpha^{(\mu)}_0 \alpha^{(\mu)}_1}$,
$\mu=0$,
1,~2, whose parameters $\alpha^{(\mu)}_0$, $\alpha^{(\mu)}_1$ take some
appropriate values corresponding to type (I.1.1) spectra. We shall actually
prove
that one may assume
\begin{equation}
  \Ap_{\mu} = \ap_{\mu}, \qquad A_{\mu} = a_{\mu}, \qquad \mu = 0, 1, 2.
  \label{eq:super-A}
\end{equation}
\par
%
%
{}For such a purpose, let us start with some algebra \algthree, and from its
generators let us construct the operators
\begin{equation}
  {\cal H}^{(\mu)} = F(N + \mu) = N + \mu I + \alpha_0 P_{1-\mu} - \alpha_2
  P_{2-\mu},     \label{eq:super-H}
\end{equation}
where in the last step we used equations~(\ref{eq:proj}), and~(\ref{eq:F}).
It is
straightforward to see that the eigenvalues ${\cal E}^{(\mu)}_n$ of~${\cal
H}^{(\mu)}$ satisfy equation~(\ref{eq:super-spec}) with $\omega_{\mu} = 1 +
\alpha_{\mu}$, $\mu=0$, 1,~2, and $\Omega_3 = 3$. For this result to be
meaningful, the conditions $\omega_{\mu} > 0$, $ \mu=0$, 1,~2, have to be
fulfilled. The latter imply the following restrictions on
$\alpha_0$,~$\alpha_1$,
\begin{equation}
  - 1 < \alpha_0 < 2, \qquad - 1 < \alpha_1 < 1 - \alpha_0.
\label{eq:super-alpha}
\end{equation}
The parameter values satisfying equation~(\ref{eq:super-alpha}) form a subset of
the set of allowed parameter values for type (I.1.1) spectra, as defined in
equation~(\ref{eq:period-spec}).\par
%
%
{}From equation~(\ref{eq:F-Fock}), it results that ${\cal H}^{(0)}$ and ${\cal
H}^{(1)}$, defined in equation~(\ref{eq:super-H}), can be rewritten as
${\cal H}^{(0)}
= \ap a$ and ${\cal H}^{(1)} = a \ap$, respectively. Comparing with
equation~(\ref{eq:hierarchy}), we conclude that equation~(\ref{eq:super-A})
is valid
for $\mu=0$, provided we define $\ap_0 \equiv \ap$, $a_0 \equiv a$, so that the
corresponding algebra parameters are $\alpha^{(0)}_0 = \alpha_0$,
$\alpha^{(0)}_1
= \alpha_1$.\par
%
%
Let us now define $\ap_1$, $a_1$, and $\ap_2$, $a_2$ in such a way that
equation~(\ref{eq:super-A}) is also valid for $\mu=1$, and $\mu=2$. From
equations~(\ref{eq:hierarchy}) and~(\ref{eq:super-H}), we obtain
\begin{eqnarray}
  {\cal H}^{(1)} & = & \ap_1 a_1 + (1 + \alpha_0) I = N + I + \alpha_0 P_0
- \alpha_2
           P_1, \nonumber \\
  {\cal H}^{(2)} & = & a_1 \ap_1 + (1 + \alpha_0) I = N + 2I - \alpha_2 P_0 +
           \alpha_0 P_2,
\end{eqnarray}
and
\begin{eqnarray}
  {\cal H}^{(2)} & = & \ap_2 a_2 + (2 + \alpha_0 + \alpha_1) I = N + 2I -
\alpha_2
           P_0 + \alpha_0 P_2, \nonumber \\
  {\cal H}^{(3)} & = & a_2 \ap_2 + (2 + \alpha_0 + \alpha_1) I = N + 3I +
\alpha_0
           P_1 - \alpha_2 P_2,
\end{eqnarray}
from which we derive
\begin{equation}
  \left[a_1, \ap_1\right] = I + \alpha_1 P_0 + \alpha_2 P_1 + \alpha_0 P_2,
\qquad
  \left[a_2, \ap_2\right] = I + \alpha_2 P_0 + \alpha_0 P_1 + \alpha_1 P_2.
\end{equation}
\par
%
%
{}Finally, from equation~(\ref{eq:super-H}), it results that ${\cal
H}^{(3)} = {\cal
H}^{(0)} + 3I$, so that $\Ap_3 = \Ap_0 = \ap_0$, $A_3 = A_0 = a_0$, as it shoud
be.\par
%
%
We conclude that the choice made in equations~(\ref{eq:super-A}),
(\ref{eq:super-H}), and~(\ref{eq:super-alpha}) provides an algebraic realization of
SSQM for any cyclic shape invariant potential of period three.\footnote{It is
obvious that by an appropriate change of energy scale, one can get any
$\Omega_3$~value instead of $\Omega_3 = 3$, as considered here.} The matrix
elements of the supersymmetric Hamiltonians and supercharges ${\cal H}_{\mu}$,
$\Qp_{\mu}$, $Q_{\mu}$, $\mu=0$, 1,~2, are expressed in terms of boson-like
operators $\ap_{\mu}$, $a_{\mu}$, $\mu=0$, 1,~2, belonging to $C_3$-extended
oscillator algebras \algthree, \algthreebis, \algthreeter, respectively, where
$\alpha_0$, $\alpha_1$ are related to the level spacings through the relations
$\omega_0 = 1 + \alpha_0$, $\omega_1 = 1 + \alpha_1$, $\omega_2 = 1 - \alpha_0 -
\alpha_1$, and restricted to those values satisfying
equation~(\ref{eq:super-alpha}). The commutators of such operators $\ap_{\mu}$,
$a_{\mu}$ are given by
\begin{equation}
  \left[a_{\mu}, \ap_{\mu}\right] = I + \alpha^{(\mu)}_0 P_0 +
\alpha^{(\mu)}_1 P_1
  + \alpha^{(\mu)}_2 P_2,
\end{equation}
where the parameters $\alpha^{(\mu)}_{\nu} \equiv \alpha_{\nu+\mu}$, $\nu=0$,
1,~2, fulfil relations similar to equation~(\ref{eq:super-alpha}), i.e.,
\begin{equation}
  - 1 < \alpha^{(\mu)}_0 < 2, \qquad - 1 < \alpha^{(\mu)}_1 < 1 -
\alpha^{(\mu)}_0.
  \label{eq:SSQM-param}
\end{equation}
For different $\mu$ values, the sets $\{ \alpha^{(\mu)}_0, \alpha^{(\mu)}_1,
\alpha^{(\mu)}_2 \}$ only differ from one another by a cyclic permutation.\par
%
%
As a final point, we would like to stress that the Hamiltonians ${\cal
H}^{(\mu)}$,
given in equation~(\ref{eq:super-H}), differ from the corresponding bosonic
oscillator Hamiltonians $H_0^{(\mu)} \equiv \half \left\{a_{\mu},
\ap_{\mu}\right\}$ through a linear combination of projection
operators~$P_{\nu}$,
\begin{equation}
  {\cal H}^{(\mu)} = H_0^{(\mu)} - \half \sum_{\nu} \left(1 +
\alpha^{(\mu)}_{\nu}
  \right)P_{\nu} + {\cal E}^{(\mu)}_0 I,
\end{equation}
or
\begin{eqnarray}
  {\cal H}^{(0)} & = & H_0^{(0)} - \half \sum_{\nu} (1 + \alpha_{\nu}) P_{\nu},
            \nonumber \\
  {\cal H}^{(1)} & = & H_0^{(1)} + \half \sum_{\nu} (1 + 2 \alpha_0 -
\alpha_{\nu+1})
            P_{\nu} \nonumber \\
  & = & H_0^{(0)} + \half \sum_{\nu} (1 + \alpha_{\nu}) P_{\nu}, \nonumber \\
  {\cal H}^{(2)} & = & H_0^{(2)} + \half \sum_{\nu} (3 - 2 \alpha_2 -
\alpha_{\nu+2})
            P_{\nu} \nonumber \\
  & = & H_0^{(0)} + \half \sum_{\nu} (3 + \alpha_{\nu+1} - \alpha_{\nu+2})
            P_{\nu}.
\end{eqnarray}
This explains why the ${\cal H}^{(\mu)}$ and $H_0^{(\mu)}$ spectra,
corresponding
to parameter values satisfying equation~(\ref{eq:SSQM-param}), consist of
periodically spaced levels characterized by different $\omega_{\nu}$ values,
although in both cases the level order is similar, and actually coincides
with that
of the standard harmonic oscillator.\par
%
%
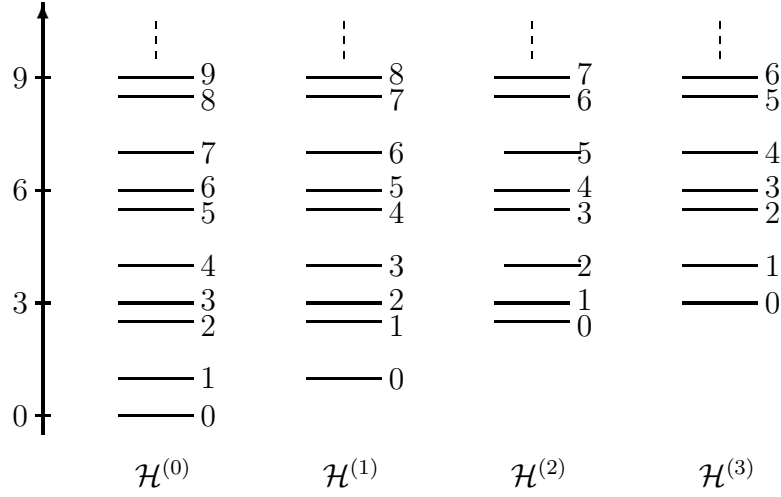
\begin{figure}

\begin{picture}(130,65)(-45,-12)

\put(-10,-2.5){\thicklines\vector(0,1){57.5}}
\multiput(-11,0)(0,15){4}{\ord}
\put(-17,0){\makebox(5,0)[r]{0}}
\put(-17,15){\makebox(5,0)[r]{3}}
\put(-17,30){\makebox(5,0)[r]{6}}
\put(-17,45){\makebox(5,0)[r]{9}}

\put(6,-7.5){\makebox(0,0){${\cal H}^{(0)}$}}

\multiput(0,0)(0,15){3}{\put(0,0){\level}\put(0,5){\level}\put(0,12.5){\level}}
\put(0,45){\level}
\put(12,0){\makebox(0,0){0}}
\put(12,5){\makebox(0,0){1}}
\put(12,12){\makebox(0,0){2}}
\put(12,15.5){\makebox(0,0){3}}
\put(12,20){\makebox(0,0){4}}
\put(12,27){\makebox(0,0){5}}
\put(12,30.5){\makebox(0,0){6}}
\put(12,35){\makebox(0,0){7}}
\put(12,42){\makebox(0,0){8}}
\put(12,45.5){\makebox(0,0){9}}
\put(5,47.5){\usebox{\dash}}

\put(31,-7.5){\makebox(0,0){${\cal H}^{(1)}$}}

\multiput(25,5)(0,15){3}{\put(0,0){\level}\put(0,7.5){\level}\put(0,10){\level}}
\put(37,5){\makebox(0,0){0}}
\put(37,12){\makebox(0,0){1}}
\put(37,15.5){\makebox(0,0){2}}
\put(37,20){\makebox(0,0){3}}
\put(37,27){\makebox(0,0){4}}
\put(37,30.5){\makebox(0,0){5}}
\put(37,35){\makebox(0,0){6}}
\put(37,42){\makebox(0,0){7}}
\put(37,45.5){\makebox(0,0){8}}
\put(30,47.5){\usebox{\dash}}

\put(56,-7.5){\makebox(0,0){${\cal H}^{(2)}$}}

\multiput(50,12.5)(0,15){2}{\put(0,0){\level}\put(0,2.5){\level}\put(0,7.5){
\level}}
\put(50,42.5){\level}
\put(50,45){\level}
\put(62,12){\makebox(0,0){0}}
\put(62,15.5){\makebox(0,0){1}}
\put(62,20){\makebox(0,0){2}}
\put(62,27){\makebox(0,0){3}}
\put(62,30.5){\makebox(0,0){4}}
\put(62,35){\makebox(0,0){5}}
\put(62,42){\makebox(0,0){6}}
\put(62,45.5){\makebox(0,0){7}}
\put(55,47.5){\usebox{\dash}}

\put(81,-7.5){\makebox(0,0){${\cal H}^{(3)}$}}

\multiput(75,15)(0,15){2}{\put(0,0){\level}\put(0,5){\level}
\put(0,12.5){\level}}
\put(75,45){\level}
\put(87,15){\makebox(0,0){0}}
\put(87,20){\makebox(0,0){1}}
\put(87,27){\makebox(0,0){2}}
\put(87,30.5){\makebox(0,0){3}}
\put(87,35){\makebox(0,0){4}}
\put(87,42){\makebox(0,0){5}}
\put(87,45.5){\makebox(0,0){6}}
\put(80,47.5){\usebox{\dash}}

\end{picture}

\caption{Spectra of the Hamiltonians ${\cal H}^{(\mu)}$, $\mu=0$, 1, 2,~3,
defined in equation~(\ref{eq:super-H}), for $\alpha_0=0$, $\alpha_1=\half$.}

\end{figure}
%
%
On figure~8 are displayed the spectra of ${\cal H}^{(\mu)}$, $\mu=0$, 1,
2,~3, for
$\alpha_0 = 0$, and $\alpha_1 = \half$. The corresponding values
of~$\omega_{\nu}$ are $\omega_0 = 1$, $\omega_1 = \frac{3}{2}$, $\omega_2 =
\half$, and the associated $C_3$-extended oscillator algebras are ${\cal
A}^{(3)}_{0,1/2}$, ${\cal A}^{(3)}_{1/2,-1/2}$, ${\cal A}^{(3)}_{-1/2,0}$,
respectively.\par
%
%
\section{Concluding Remarks}   \label{sec:conclusion}
In the present paper, we considered a bosonic oscillator Hamiltonian~$H_0$,
associated with the $C_3$-extended oscillator algebra \algthree\ introduced
in~\cite{cq98a}, and we studied its spectrum in terms of the algebra parameters
$\alpha_0$,~$\alpha_1$. We showed that such a spectrum has a very rich
structure,
contrary to what happens for the two-particle Calogero Hamiltonian, connected
with the $C_2$ (or $S_2$)-extended oscillator algebra \algtwo\ (also referred to
as the Calogero-Vasiliev algebra). In particular, we obtained both nondegenerate
spectra, with or without the same level order as the standard harmonic
oscillator,
and spectra exhibiting some double and/or triple degeneracies.\par
%
%
More importantly, we pointed out that some of the nondegenerate spectra, namely
those of type (I.1.1), (II.1.1.1), and (III.1.1.1), have an infinite number
of periodically
spaced levels, as the spectra arising in SSQM when considering cyclic shape
invariant potentials of period three~\cite{sukhatme}. We finally obtained a
matrix
realization of the supersymmetric Hamiltonians and supercharges associated with
the latter in terms of creation and annihilation operators $\ap_{\mu}$,
$a_{\mu}$,
$\mu=0$, 1,~2, belonging to $C_3$-extended oscillator algebras, whose parameters
are obtained by cyclic permutations from a starting set $\{\alpha_0, \alpha_1,
\alpha_2\}$, for which $-1 < \alpha_0 <2$, $-1 < \alpha_1 < 1 - \alpha_0$, and
$\alpha_2 = - \alpha_0 - \alpha_1$.\par
%
%
It is obvious that the results derived in the present paper can be extended to
bosonic oscillator Hamiltonians~$H_0$ associated with $C_{\lambda}$-extended
oscillator algebras \alg, corresponding to $\lambda$~values different from
three.
Although the complete classification of their possible types of spectra in
terms of
the algebra parameters $\alpha_0$, $\alpha_1$,~$\ldots$, $\alpha_{\lambda-2}$,
becomes rather complicated for $\lambda > 3$, generalizing the results for
spectra with periodically spaced levels is straightforward. In particular,
it can
easily be shown that the hierarchy of supersymmetric Hamiltonians and
supercharges $\{\, {\cal H}_{\mu}, \Qp_{\mu}, Q_{\mu} \mid \mu = 0, 1, \ldots,
\lambda-1 \,\}$ of~\cite{sukhatme}, corresponding to cyclic shape invariant
potentials of period~$\lambda \ge 2$, can be built from creation and
annihilation
operators $\ap_{\mu}$, $a_{\mu}$, $\mu=0$, 1, $\ldots$,~$\lambda-1$,
belonging to
$C_{\lambda}$-extended oscillator algebras, whose parameters are obtained by
cyclic permutations from a starting set $\{\alpha_0, \alpha_1, \ldots,
\alpha_{\lambda-1}\}$, for which $-1 < \alpha_0 < \lambda-1$, $-1 <
\alpha_{\mu} < \lambda - \mu -1 - \sum_{\nu=0}^{\mu-1} \alpha_{\nu}$ if
$\mu=1$, 2, $\ldots$,~$\lambda-2$, and $\alpha_{\lambda-1} = -
\sum_{\nu=0}^{\lambda-2} \alpha_{\nu}$.\par
%
%
A very interesting open question is the possibility of realizing
$C_{\lambda}$-extended oscillator algebras in terms of differential operators.
Since one-dimensional Hamiltonians are known to have no degeneracies in their
bound state spectrum, the existence of degeneracies in the $H_0$ spectrum for
some parameter values shows that such a realization should at least involve
two variables.\par
%
%


\begin{thebibliography}{99}

\bibitem{gendenshtein} L.E. Gendenshtein, JETP Lett. {\bf 38} (1983) 356.

\bibitem{witten} E. Witten, Nucl. Phys. B {\bf 185} (1981) 513.

\bibitem{cooper} F. Cooper, A. Khare and U. Sukhatme, Phys. Rep. {\bf 251}
(1995)
267.

\bibitem{sukhatme} U.P. Sukhatme, C. Rasinariu and A. Khare, Phys. Lett. A
{\bf 234}
(1997) 401.

\bibitem{gango} A. Gangopadhyaya and U.P. Sukhatme, Phys. Lett. A {\bf 224}
(1996)
5.

\bibitem{green} H.S. Green, Phys. Rev. {\bf 90} (1953) 270; \\
Y. Ohnuki and S. Kamefuchi, {\em Quantum Field Theory and Parastatistics}
(Springer, Berlin, 1982).

\bibitem{beckers95} J. Beckers, N. Debergh and A.G. Nikitin, Fortschr.
Phys. {\bf
43} (1995) 67, 81; \\
J. Beckers and N. Debergh, Int. J. Mod. Phys. A {\bf 10} (1995) 2783.

\bibitem{mishra} A.K. Mishra and G. Rajasekaran, Pramana (J. Phys.) {\bf
36} (1991)
537, {\bf 37} (1991) 455(E).

\bibitem{rubakov} V.A. Rubakov and V.P. Spiridonov, Mod. Phys. Lett. A {\bf
3} (1988)
1337;\\
A. Khare, J. Math. Phys. {\bf 34} (1993) 1277.

\bibitem{beckers90} J. Beckers and N. Debergh, Nucl. Phys. B {\bf 340}
(1990) 767.

\bibitem{khare} A. Khare, A.K. Mishra and G. Rajasekaran, Int. J. Mod.
Phys. A {\bf 8}
(1993) 1245.

\bibitem{drinfeld} V.G. Drinfeld, in {\em Proc. Int. Congr. of Mathematicians
(Berkeley, CA)}, ed. A.M.~Gleason (American Mathematical Society,
Providence, RI,
1986) p.~798; \\
M. Jimbo, Lett. Math. Phys. {\bf 10} (1985) 63, {\bf 11} (1986) 247.

\bibitem{arik} M. Arik and D.D. Coon, J. Math. Phys. {\bf 17} (1976) 524.

\bibitem{biedenharn} L.C. Biedenharn, J. Phys. A {\bf 22} (1989) L873;\\
A.J. Macfarlane, J. Phys. A {\bf 22} (1989) 4581; \\
C.-P. Sun and H.-C. Fu, J. Phys. A {\bf 22} (1989) L983.

\bibitem{jannussis} A. Jannussis, G. Brodimas and R. Mignani, J. Phys. A
{\bf 24}
(1991) L775; \\
A. Jannussis, J. Phys. A {\bf 26} (1993) L233.

\bibitem{daska91} C. Daskaloyannis, J. Phys. A {\bf 24} (1991) L789.

\bibitem{irac} M. Irac-Astaud and G. Rideau, On the existence of quantum
bihamiltonian systems: The harmonic oscillator case, Universit\'e Paris VII
preprint, PAR-LPTM92; Lett. Math. Phys. {\bf 29} (1993) 197; Theor. Math. Phys.
{\bf 99} (1994) 658.

\bibitem{mcdermott} R.J. McDermott and A.I. Solomon, J. Phys. A {\bf 27} (1994)
L15.

\bibitem{meljanac} S. Meljanac, M. Milekovi\'c and S. Pallua, Phys. Lett. B
{\bf 328}
(1994) 55; \\
S. Meljanac and M. Milekovi\'c, Int. J. Mod. Phys. A {\bf 11} (1996) 1391.

\bibitem{katriel} J. Katriel and C. Quesne, J. Math. Phys. {\bf 37} (1996) 1650.

\bibitem{cq95b} C. Quesne and N. Vansteenkiste, J. Phys. A {\bf 28} (1995)
7019.

\bibitem{cq96} C. Quesne and N. Vansteenkiste, Helv. Phys. Acta {\bf 69} (1996)
141; Czech. J. Phys. {\bf 47} (1997) 115.

\bibitem{daska92} C. Daskaloyannis, J. Phys. A {\bf 25} (1992) 2261;\\
D. Bonatsos and C. Daskaloyannis, Chem. Phys. Lett. {\bf 203} (1993) 150.

\bibitem{cq94a} C. Quesne, Phys. Lett. A {\bf 193} (1994) 245.

\bibitem{bonatsos93a} D. Bonatsos, C. Daskaloyannis and K. Kokkotas, Phys.
Rev. A
{\bf 48} (1993) R3407, {\bf 50} (1994) 3700.

\bibitem{greenberg} O.W. Greenberg, Phys. Rev. Lett. {\bf 64} (1990) 705; Phys.
Rev. D {\bf 43} (1991) 4111; \\
D.I. Fivel, Phys. Rev. Lett. {\bf 65} (1990) 3361.

\bibitem{chaturvedi} S. Chaturvedi and V. Srinivasan, Phys. Rev. A {\bf 44}
(1991)
8024;\\
A.J. Macfarlane, J. Math. Phys. {\bf 35} (1994) 1054.

\bibitem{cq94b} C. Quesne, J. Phys. A {\bf 27} (1994) 5919.

\bibitem{vasiliev} M.A. Vasiliev, Int. J. Mod. Phys. A {\bf 6} (1991) 1115.

\bibitem{poly} A.P. Polychronakos, Phys. Rev. Lett. {\bf 69} (1992) 703; \\
L. Brink, T.H. Hansson and M.A. Vasiliev, Phys. Lett. B {\bf 286} (1992) 109;\\
L. Brink and M.A. Vasiliev, Mod. Phys. Lett. A {\bf 8} (1993) 3585.

\bibitem{brze} T. Brzezi\'nski, I.L. Egusquiza and A.J. Macfarlane, Phys.
Lett. B {\bf
311} (1993) 202.

\bibitem{calogero} F. Calogero, J.~Math.~Phys. {\bf 10} (1969) 2191, 2197,
{\bf 12}
(1971) 419.

\bibitem{cq95a} C. Quesne, Mod. Phys. Lett. A {\bf 10} (1995) 1323.

\bibitem{bonatsos93b} D. Bonatsos and C. Daskaloyannis, Phys. Lett. B {\bf 307}
(1993) 100.

\bibitem{plyu} M.S. Plyushchay, Mod. Phys. Lett. A {\bf 11} (1996) 397;
Ann. Phys.
(NY) {\bf 245} (1996) 339.

\bibitem{beckers97} J. Beckers, N. Debergh and A.G. Nikitin, Int. J. Theor.
Phys. {\bf
36} (1997) 1991.

\bibitem{cq98a} C. Quesne and N. Vansteenkiste, Phys. Lett. A {\bf 240} (1998)
21.

\bibitem{cq98b} C. Quesne and N. Vansteenkiste, $C_{\lambda}$-extended
oscillator algebras and some of their deformations (in preparation).

\bibitem{cornwell} J.F. Cornwell, {\em Group Theory in Physics}
(Academic, New York, 1984) vol.~1, p.~117.

\end {thebibliography}
\end{document}